\documentclass[11pt]{article}
\usepackage[preprint]{acl}
\usepackage{times}
\usepackage{latexsym}
\usepackage[T1]{fontenc}
\usepackage[utf8]{inputenc}
\usepackage{microtype}
\usepackage{pifont}
\usepackage{enumitem}
\usepackage{inconsolata}
\usepackage{adjustbox}
\usepackage[hang,flushmargin]{footmisc}
\usepackage{graphicx}
\usepackage{subcaption}
\usepackage{booktabs}
\usepackage{amsmath}
\usepackage{amssymb}
\usepackage{float}
\usepackage[normalem]{ulem}

\usepackage{bm}
\usepackage{multirow}
\usepackage[table]{xcolor}

\title{Skim and Skip: Hierarchical Adaptive Inference for Efficient Multimodal Retrieval}

\author{
\textbf{Meng Gao}$^{1,2,*}$ \quad
\textbf{Yizhen Zhang}$^{1,2,*}$ \quad
\textbf{Yang Ding}$^{1,*}$ \quad
\textbf{Ziqi Dai} \quad
\textbf{Shuoshuo Zhang}$^{1}$ \\
\textbf{Junjie Wang}$^{1}$ \quad
\textbf{Taiqiang Wu}$^{1}$ \quad
\textbf{Chufan Shi}$^{1}$ \quad
\textbf{Lei Ji}$^{2,\dagger}$ \quad
\textbf{Jian Jiao}$^{2}$ \\
\textbf{Linfeng Zhang}$^{3,\dagger}$ \quad
\textbf{Yeyun Gong}$^{2}$ \quad
\textbf{Yujiu Yang}$^{1,\dagger}$ \\
$^{1}$Tsinghua University \quad
$^{2}$Microsoft \quad
$^{3}$Shanghai Jiao Tong University
}

\begin{document}
\maketitle

\begingroup
\renewcommand\thefootnote{}
\footnotetext{$^*$Equal contribution. Meng Gao and Yizhen Zhang did this work during the internship at Microsoft Research Asia.\\
$^\dagger$Corresponding authors.\\
Under review.}
\endgroup

\begin{abstract}
Universal multimodal retrieval (UMR) increasingly adopts multimodal large language models (MLLMs) as unified embedding backbones, but their strong retrieval performance comes at substantial inference cost. Existing methods typically rely on uniformly dense inference, where all input tokens are processed through the entire model and matched using the final-layer [EOS] representation. However, this paradigm overlooks two key forms of heterogeneity in multimodal retrieval: token contributions to the final retrieval embedding are highly uneven, and different queries require markedly different amounts of inference depth. To address this, we propose \textbf{Skim and Skip (SAS)}, a hierarchical adaptive inference framework for efficient multimodal retrieval. SAS first performs token-level evidence selection to preserve only the input information most relevant to the final retrieval embedding, and then performs depth-adaptive inference to determine whether the current representation is already sufficient for reliable matching. Experiments on 12 MMEB retrieval tasks show that SAS retains about 99\% of the dense baseline's average retrieval performance while achieving up to \textbf{1.64$\times$} end-to-end speedup and up to \textbf{66.3\%} FLOPs reduction. Our code is available at \url{https://github.com/GaoMengGladys/SAS}.
\end{abstract}

\section{Introduction}

\begin{figure}[t]
    \centering
    \includegraphics[width=\linewidth]{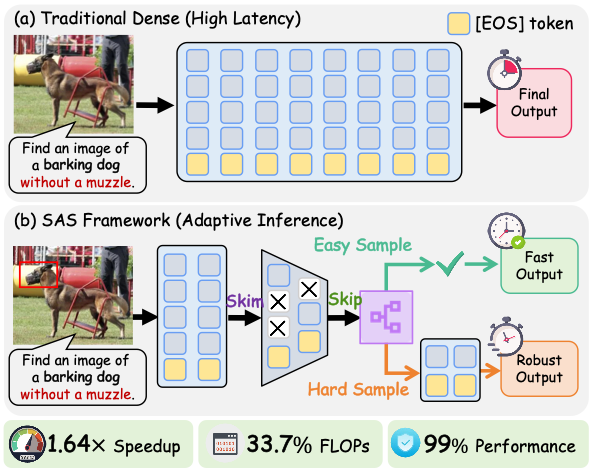}
    \caption{\textbf{Motivation of Skim and Skip (SAS).}
    Conventional MLLM retrieval applies uniformly dense inference to all tokens across all layers, resulting in high latency.
    In contrast, SAS performs hierarchical adaptive inference: it first filters retrieval-relevant information (\emph{Skim}) and then allocates depth based on representation maturity (\emph{Skip}), allowing easy queries to exit early while reserving deeper computation for harder ones.}
    \label{fig:teaser-motivation}
\end{figure}

Multimodal large language models (MLLMs) are increasingly used as unified backbones for multimodal retrieval~\citep{Jiang2024VLM2VecTV,Zhang2025BridgingMI,he2026plume}. Benefiting from large-scale cross-modal pretraining, MLLM-based retrievers have achieved strong performance across zero-shot, fine-grained, and universal retrieval settings~\citep{Liu2024LamRALM,Jiang2024VLM2VecTV,he2026plume,li2026magic,huang2026mmeb, Yang2024VisionZipLI, zhang2025generative, zhang2026omniverifier, lin2024rho}. However, this performance comes at substantial inference cost, as existing methods typically process every query with all tokens through the full depth of the model~\citep{lyu2025puma,xiao2025metaembed}.

The high inference cost primarily arises from two core computational factors: input information width and information processing depth. First, multimodal inputs—comprising visual and textual tokens—are long and processed uniformly, despite being inherently uneven, redundant, and noisy, which degrades the quality of the final retrieval embedding. Second, different queries require varying levels of processing depth, yet existing approaches often apply a fixed processing budget. Therefore, efficient retrieval hinges on two tightly coupled questions: which portions of the input are truly informative for constructing the representation, and how much computation is necessary for that representation to become sufficiently reliable.

Existing acceleration methods typically address these two issues in isolation, either compressing the input sequence~\citep{Yang2024VisionZipLI,Chen2024AnII,li2026magic,wu2026hidrop,deniz2025adsc,he2026energy}, or executing fewer layers~\citep{lyu2025puma,yoo2026adept,huang2024raee,yang2026dysl}. However, standard pruning strategies developed for multimodal tasks such as VQA are poorly suited to retrieval: existing token-reduction methods largely overlook the critical role of the EOS embedding, causing severe performance degradation even at low compression ratios. Moreover, naive depth reduction also harms performance: while easy queries already achieve 55.0\% Hit@1 at intermediate layer 22, hard queries reach only 7.2\% on MSCOCO, indicating substantial variability in required processing depth. Furthermore, jointly learning what to preserve and when a representation suffices is inherently challenging, as it deviates from standard full-token supervision.

Motivated by this view, we propose \textbf{Skim and Skip (SAS)}, a hierarchical adaptive inference framework for efficient multimodal retrieval. SAS consists of two stages. \textbf{Skim} performs \texttt{[EOS]}-centric cross-modal information selection, preserving only the input information most relevant to the final retrieval embedding. \textbf{Skip} performs adaptive depth allocation, deciding whether the current representation is already mature enough for reliable retrieval. To motivate this design, we revisit the assumptions behind uniformly dense inference in Section~\ref{sec:analysis}, and then present SAS in Section~\ref{sec:method}.

Our main contributions are as follows:
\begin{itemize}[wide=0pt, itemsep=0em, topsep=0em]
    \item We identify two key forms of heterogeneity overlooked by uniform dense inference in MLLM retrieval: uneven token contributions to the final retrieval representation, and varying depth requirements across queries.
    \item We propose \textbf{Skim and Skip (SAS)}, a hierarchical adaptive inference framework that unifies \texttt{[EOS]}-centric cross-modal information selection and adaptive depth allocation for multimodal retrieval.
    \item We evaluate SAS on 12 multimodal retrieval benchmarks spanning five task types. SAS retains about 99\% of the dense baseline's average retrieval performance while achieving up to 1.64$\times$ end-to-end speedup and reducing FLOPs by up to 66.3\%.
\end{itemize}

\section{Revisiting Uniform Dense Inference in MLLM Retrieval}
\label{sec:analysis}

In MLLM-based retrieval, most approaches follow a uniformly dense inference paradigm, processing every query with the full set of input tokens through the entire model. While effective, this design implicitly assumes that all input information warrants equal computational effort and that all queries require the same depth of reasoning. To revisit this paradigm, we examine it from two perspectives: whether all tokens contribute equally to the final \texttt{[EOS]} retrieval embedding, and whether all queries require the same inference depth to form a reliable retrieval representation.

\subsection{Are All Tokens Equally Informative?}
\label{sec:token_analysis}

We first examine how the \texttt{[EOS]} token aggregates retrieval evidence. Specifically, we analyze the attention from \texttt{[EOS]} to all input tokens across layers of VLM2Vec-V2.0. As shown in Table~\ref{tab:attn_avg}, at the middle layer (L14), where cross-modal interaction is strongest, the top-10 tokens capture 77.8\% of the total \texttt{[EOS]} attention mass and the top-20 capture 87.9\%. These results suggest that the final \texttt{[EOS]} embedding is driven primarily by a small subset of tokens, while the majority contribute little.

\begin{table}[t]
\centering
\small
\setlength{\tabcolsep}{5pt}
\renewcommand{\arraystretch}{1.05}
\resizebox{\columnwidth}{!}{
\begin{tabular}{lccc}
\toprule
\textbf{Top-k} 
& \textbf{Shallow (L0)}
& \textbf{Middle (L14)}
& \textbf{Deep (L27)} \\
\midrule
Top-10 & 69.6\% & \textbf{77.8\%} & 58.3\% \\
Top-20 & 86.3\% & \textbf{87.9\%} & 75.1\% \\
\bottomrule
\end{tabular}
}
\caption{
Average \texttt{[EOS]} attention on 11 MMEB tasks.
}
\label{tab:attn_avg}
\end{table}
\begin{table}[t]
\centering
\small
\setlength{\tabcolsep}{5pt}
\renewcommand{\arraystretch}{1.05}
\resizebox{\columnwidth}{!}{
\begin{tabular}{lcccc}
\toprule
\textbf{Drop Ratio}
& \textbf{CIRR} 
& \textbf{FashionIQ} 
& \textbf{NIGHTS} 
& \textbf{OVEN}  \\
\midrule
0\%  & 0.572 & 0.197 & 0.685 & 0.645 \\
10\% & 0.591 & 0.204 & 0.659 & 0.647 \\
20\% & 0.577 & 0.191 & 0.667 & 0.648 \\
30\% & 0.586 & 0.194 & 0.660 & 0.633 \\
\bottomrule
\end{tabular}
}
\caption{
Negligible impact of random drops.}
\label{tab:drop_ratio}
\end{table}

\begin{figure*}[t]
    \centering
    \begin{subfigure}{0.48\linewidth}
        \centering
        \includegraphics[width=\linewidth]{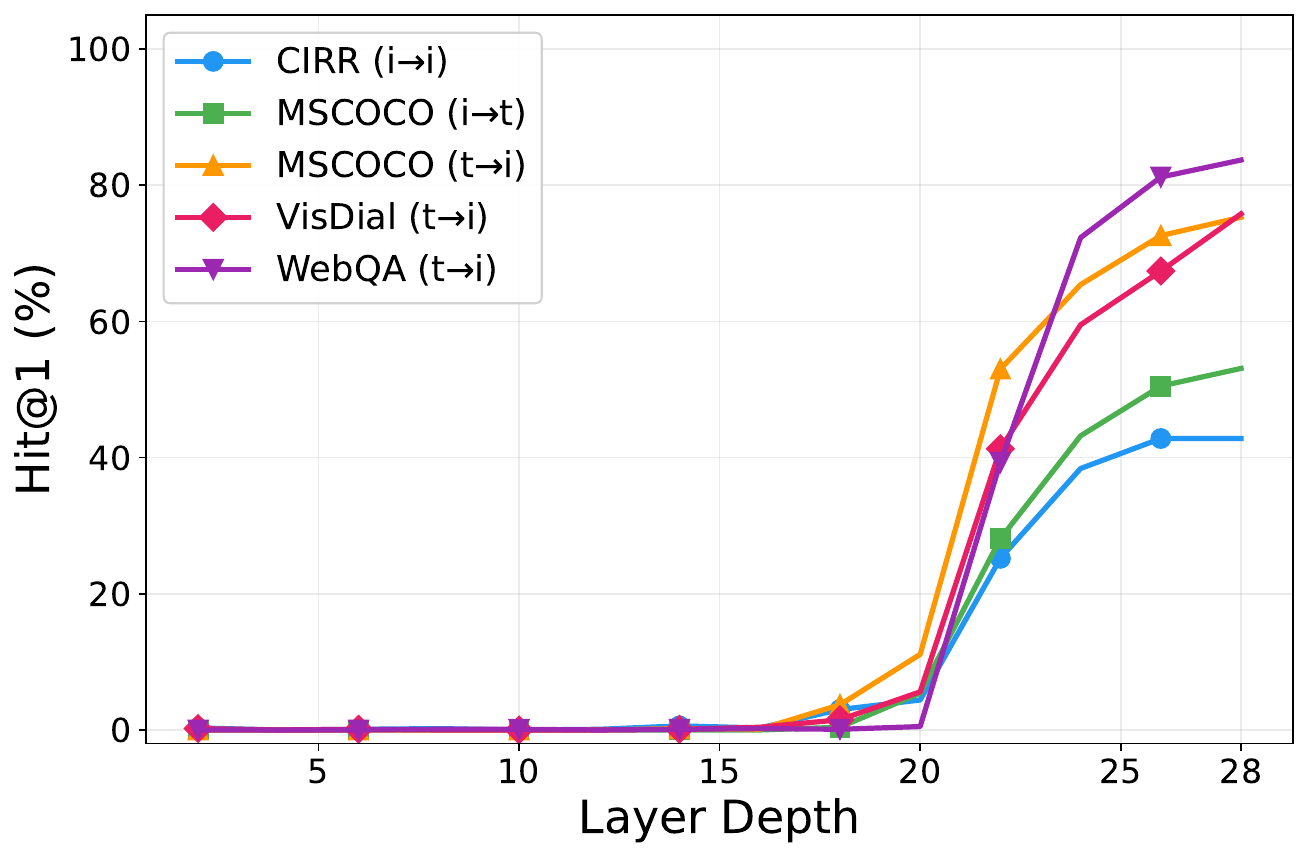}
        \caption{Per-layer retrieval performance across five tasks.}
        \label{fig:depth}
    \end{subfigure}
    \hfill
    \begin{subfigure}{0.48\linewidth}
        \centering
        \includegraphics[width=\linewidth]{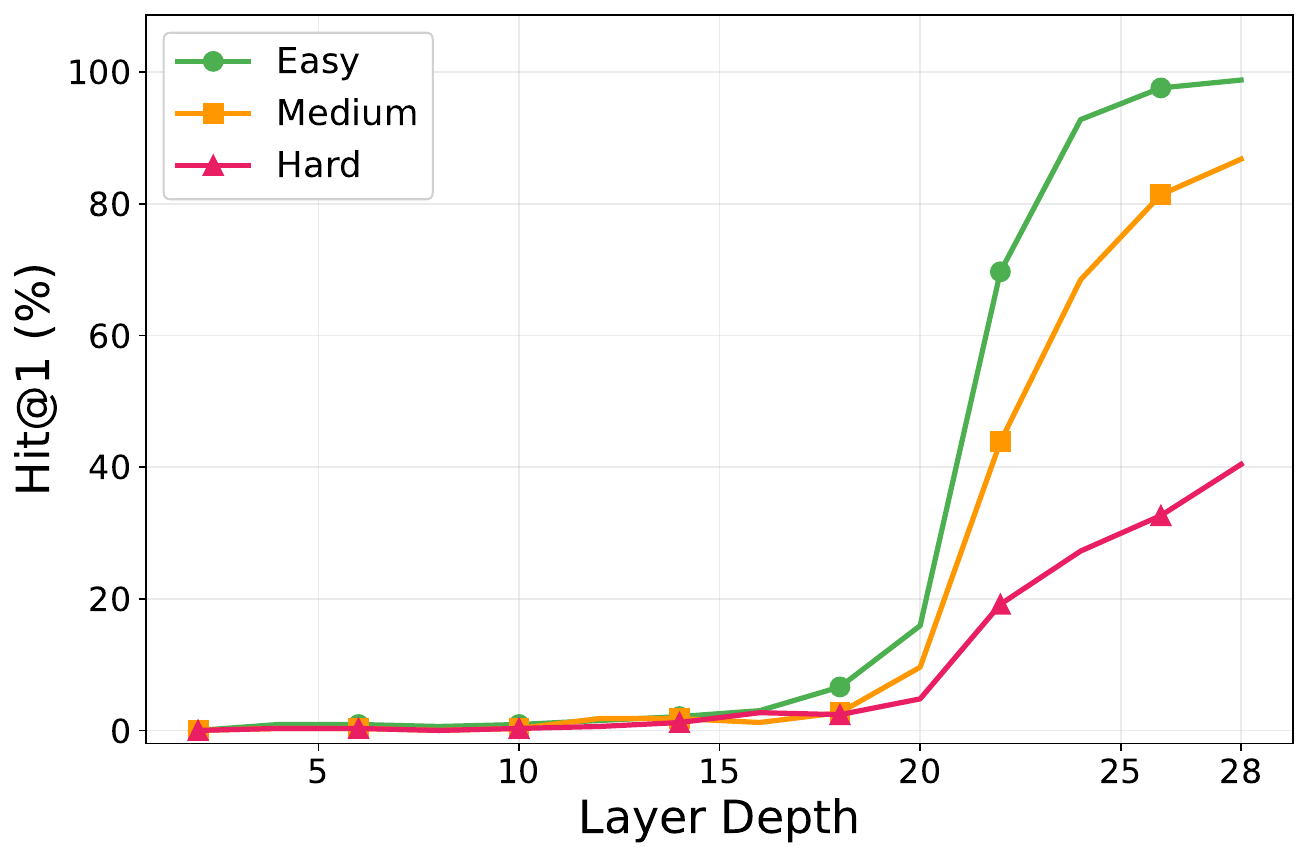}
        \caption{Easy/Medium/Hard stratification (MSCOCO t$\to$i).}
        \label{fig:stratified}
    \end{subfigure}
    \caption{Evidence against uniform depth allocation.} 
    \label{fig:depth_stratified}
\end{figure*}

This view is reinforced by random token removal experiments. We randomly drop different proportions of tokens across multiple retrieval tasks (Table~\ref{tab:drop_ratio}) and find that performance remains largely stable under 10--30\% pruning, with slight gains on some tasks. This suggests that retrieval evidence is inherently sparse and that some tokens may even introduce distracting or irrelevant signals. Efficient retrieval inference should therefore identify the information that materially contributes to the final \texttt{[EOS]} retrieval embedding and preserve only that evidence for subsequent reasoning.

Importantly, this token selection problem cannot be reduced to a modality-specific heuristic. Token composition varies substantially across tasks. Such variability makes fixed modality-dependent rules brittle. Instead, retrieval-oriented token selection should rely on a unified cross-modal criterion that directly measures a token's relevance to the final \texttt{[EOS]} representation.

\subsection{Do All Queries Require the Same Inference Depth?}
\label{sec:depth_analysis}

We next examine whether all queries require the same inference depth. Layer-wise probing on VLM2Vec-V2.0 shows that retrieval signals do not emerge only at the final layer, but develop progressively in intermediate and late layers (Figure~\ref{fig:depth_stratified}a). For example, on VisDial, the model already reaches 50.9\% Hit@1 at layer 22, compared to 82.1\% at the final layer, indicating that intermediate layers already encode substantial retrieval-relevant semantics before full-depth inference is completed.

However, the benefit of deeper computation is far from uniform across queries. To examine this heterogeneity, we partition queries into Easy, Medium, and Hard groups according to the final-layer Top-1/Top-2 similarity margin, and compare their layer-wise retrieval performance (Figure~\ref{fig:depth_stratified}b). On MSCOCO text-to-image retrieval, Easy queries already achieve 55.0\% Hit@1 at layer 22, whereas Hard queries reach only 7.2\%. This disparity shows that retrieval representations mature at very different rates across samples.

At the same time, the presence of useful intermediate-layer signals does not by itself yield a practical adaptive-depth mechanism. Standard retrieval backbones are supervised only at the final layer, so intermediate representations, although informative, are not explicitly trained to function as standalone retrieval embeddings. Converting this latent depth redundancy into real inference savings therefore requires two steps: first, activating retrieval capability at a target intermediate layer through additional supervision; and second, deciding whether inference can terminate there based on the maturity of the current representation.

\section{Methodology}
\label{sec:method}

\begin{figure*}[t]
    \centering
    \includegraphics[width=\textwidth]{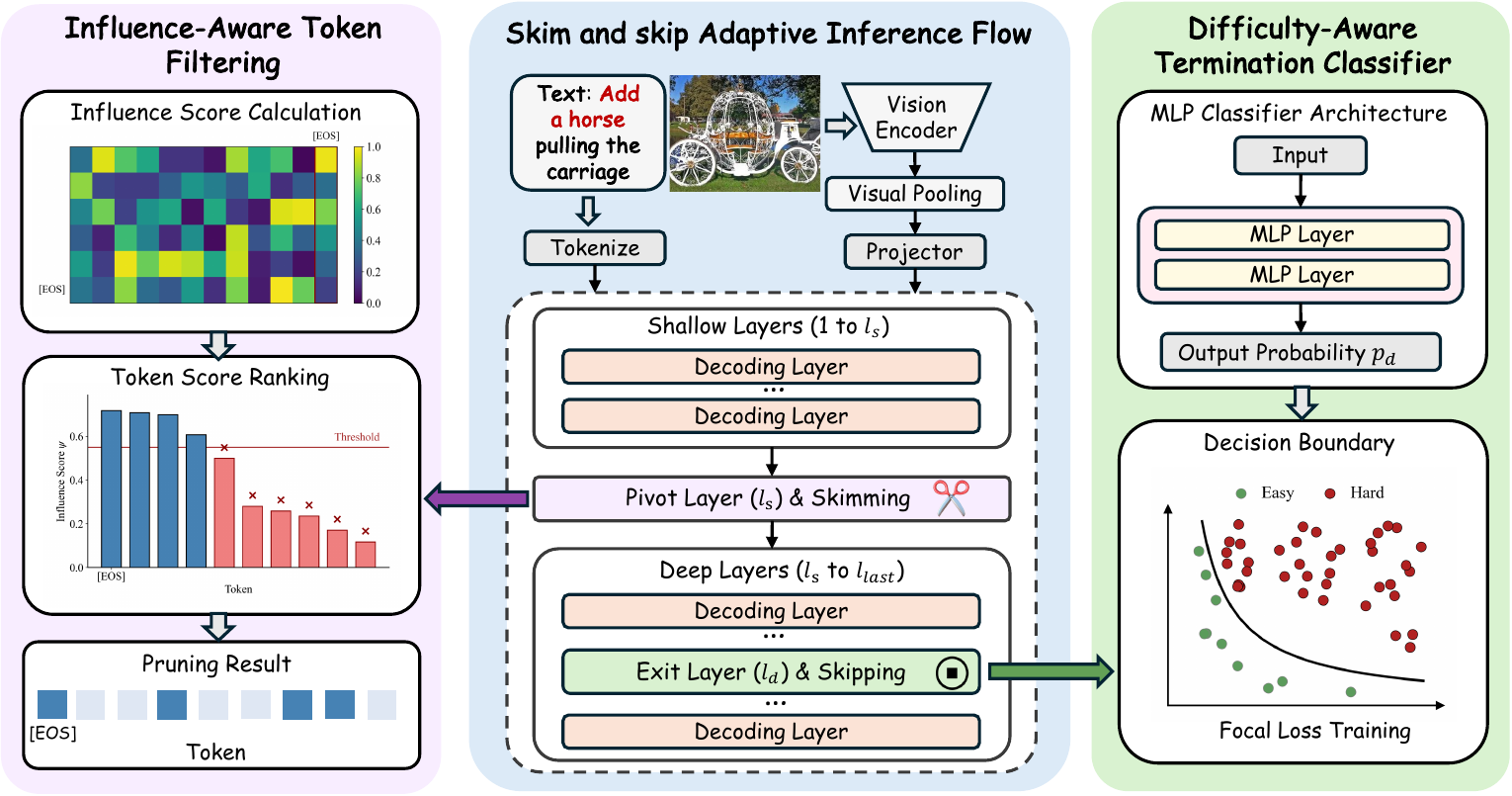}
    \caption{\textbf{Overview of SAS.} SAS treats retrieval inference as progressive evidence accumulation. \textbf{Left: Token-level evidence selection.} At the skim layer $l_s$, the Influence-Aware Filter estimates each token's marginal influence on the \texttt{[EOS]} embedding and removes low-contribution tokens, preserving only evidence-bearing information. \textbf{Right: Layer-level evidence sufficiency.} At the decision layer $l_d$, a learned sufficiency classifier determines whether the accumulated evidence is adequate for reliable retrieval; sufficiently mature queries exit early, while the rest continue to deeper layers.}
    \label{fig:main_pipeline}
\end{figure*}

The analysis in Section~\ref{sec:analysis} shows that efficient multimodal retrieval requires two coupled decisions: what information to preserve and how much computation to allocate. Motivated by this view, we propose \textbf{Skim and Skip (SAS)}, a hierarchical adaptive inference framework for multimodal retrieval. For each query, SAS first applies \textbf{Skim} to retain the input information most relevant to the final \texttt{[EOS]} retrieval embedding, and then applies \textbf{Skip} to determine whether the current representation is already mature enough for early termination. Accordingly, Skim addresses \emph{information selection}, while Skip addresses \emph{depth allocation}.

\subsection{Skim: \texttt{[EOS]}-centric Cross-modal Information Selection}

The goal of \textbf{Skim} is to retain only the input information that materially contributes to the final \texttt{[EOS]} retrieval embedding, so that subsequent computation is concentrated on informative tokens. Because visual and textual token composition varies substantially across retrieval tasks, token selection must be defined under a unified cross-modal criterion. Rather than relying on local saliency or heuristic thresholds, Skim measures token importance by the perturbation that removing a token would induce on the \texttt{[EOS]} representation.

In practice, we first apply pooling to the ViT-encoded visual tokens to shorten the initial visual sequence before it is fed into the LLM. We then perform cross-modal token selection at layer $l_s$.

\paragraph{Local Perturbation of Token Removal.}
Consider attention head $h$ at layer $l$. Let $o_{l,h}(e)$ denote the attention output at the \texttt{[EOS]} position, and let $\alpha_{l,h}(e \!\to\! j)$ denote the attention weight from \texttt{[EOS]} to token $j$. The output at \texttt{[EOS]} is
\begin{equation}
o_{l,h}(e)=\sum_i \alpha_{l,h}(e \to i)\, v_{l,h,i},
\end{equation}
where $v_{l,h,i}$ is the value vector of token $i$.

If token $j$ is removed, the attention weights over the remaining tokens are re-normalized, perturbing the head output at \texttt{[EOS]}. The resulting exact perturbation is
\begin{equation}
\Delta o_{l,h}(e)
=
\frac{\alpha_{l,h}(e \!\to\! j)}{1-\alpha_{l,h}(e \!\to\! j)}
\Big( o_{l,h}(e)-v_{l,h,j} \Big).
\label{eq:local_shift}
\end{equation}

This expression captures the immediate effect of removing a token on the \texttt{[EOS]} representation in the current layer. Importantly, the perturbation depends not only on the attention mass assigned to token $j$, but also on the discrepancy between its value vector and the aggregated head output. Attention alone is therefore insufficient to characterize a token's contribution to retrieval.

\paragraph{A Computable Upper Bound on Token Influence.}
The local perturbation first affects a single attention head and then propagates through the output projection $W_O^{l,h}$ to subsequent layers. To quantify its downstream impact, we consider the magnitude of the projected perturbation, $\|W_O^{l,h}\Delta o_{l,h}(e)\|_2$.

Computing this quantity exactly during inference is inefficient. Instead, we upper-bound it using the spectral norm $\sigma_O^{l,h}$ of $W_O^{l,h}$:
\begin{equation}
\begin{aligned}
\|W_O^{l,h}\Delta o_{l,h}(e)\|_2
&\le
\sigma_O^{l,h}
\frac{\alpha_j}{1-\alpha_j}
\\
&\quad \times
\Big(
\|o_{l,h}(e)\|_2
+
\|v_{l,h,j}\|_2
\Big),
\end{aligned}
\label{eq:bound_transition}
\end{equation}
where $\alpha_j$ abbreviates $\alpha_{l,h}(e \!\to\! j)$.

To obtain an inference-time computable score, we further bound $\|o_{l,h}(e)\|_2$ and $\|v_{l,h,j}\|_2$ using the spectral norm $\sigma_V^{l,h}$ of the value projection $W_V^{l,h}$ and its bias term $\mathbf{b}_V^{l,h}$. Aggregating the resulting bound over all $H$ heads yields the following influence score for token $j$ at layer $l$:
\begin{equation}
\begin{aligned}
\psi_l(j)
&=
\sum_{h=1}^{H}
\sigma_O^{l,h}
\frac{\alpha_j}{1-\alpha_j} \\
&\quad \cdot
\Big[
\sigma_V^{l,h}
\big(
S_{l,h}+\|\mathbf{h}_l[j]\|_2
\big)
+2\|\mathbf{b}_V^{l,h}\|_2
\Big],
\end{aligned}
\label{eq:psi_main}
\end{equation}
where
\begin{equation}
S_{l,h}:=\sum_i \alpha_{l,h}(e \!\to\! i)\,\|\mathbf{h}_l[i]\|_2,
\end{equation}
and $\mathbf{h}_l[j]$ is the input hidden state of token $j$ at layer $l$. The full derivation is provided in Appendix~\ref{sec:appendix_theory}.

This score can be interpreted as a computable upper bound on the perturbation that removing token $j$ may induce on the subsequent evolution of the \texttt{[EOS]} representation. A larger score indicates greater importance to the final retrieval embedding, while a smaller score suggests that the token can be removed with limited effect.

\paragraph{Unified Cross-modal Selection.}
At skim layer $l_s$, we rank all tokens by their influence scores and remove those with the lowest values, preserving the tokens most important to the final \texttt{[EOS]} retrieval embedding. Because the score is defined directly by a token's effect on \texttt{[EOS]}, visual and textual tokens can be ranked under the same criterion.

Moreover, the influence score is computed from attention weights and hidden states already available during the forward pass, introducing negligible overhead. 

\subsection{Skip: Adaptive Depth Allocation by Representation Maturity}

The goal of \textbf{Skip} is to determine whether the current depth is already sufficient for reliable retrieval and to allocate additional computation only when necessary. This is non-trivial because standard retrieval backbones are supervised only on the final-layer \texttt{[EOS]} embedding. As a result, intermediate layers may already contain retrieval signals, but are not explicitly trained to serve as standalone retrieval embeddings.

To address this, Skip adopts a two-stage training paradigm: we first activate retrieval capability at the target decision layer, and then train a lightweight assessor to decide whether inference can stop there.

\paragraph{Stage I: Activating Intermediate Retrieval Capability.}
To equip the target layer $l_d$ with standalone retrieval capability, we supervise its representation with both feature-level distillation and contrastive learning. Intuitively, this stage encourages the intermediate embedding to both align with the final-layer semantic space and remain discriminative for retrieval.

Specifically, we treat the final-layer embedding $e^{(L)}$ as the teacher and the embedding at layer $l_d$, denoted by $e^{(l_d)}$, as the student. We first align them using mean squared error:
\begin{equation}
\mathcal{L}_{distill}
=
\mathbb{E}_{(q,c)\sim\mathcal{D}}
\Big[
\|e_q^{(L)}-e_q^{(l_d)}\|_2^2
+
\|e_c^{(L)}-e_c^{(l_d)}\|_2^2
\Big].
\end{equation}
This term reduces the representation gap between intermediate and final embeddings.

To ensure retrieval discriminability, we further apply InfoNCE supervision~\citep{infoNCE} to the intermediate embedding:
\begin{equation}
\mathcal{L}_{con}^{(l_d)}
=
-\frac{1}{N}
\sum_{i=1}^{N}
\log
\frac{
\exp(\mathrm{sim}(e_{q_i}^{(l_d)}, e_{c_i}^{(l_d)})/\tau)
}{
\sum_{j=1}^{N}
\exp(\mathrm{sim}(e_{q_i}^{(l_d)}, e_{c_j}^{(l_d)})/\tau)
},
\end{equation}
where $\mathrm{sim}(\cdot,\cdot)$ denotes cosine similarity and $\tau$ is a temperature parameter.

We combine the two objectives with a dynamic weight:
\begin{equation}
\mathcal{L}_{inter}(t)
=
\lambda(t)\mathcal{L}_{con}^{(l_d)}
+
(1-\lambda(t))\mathcal{L}_{distill},
\label{eq:inter_loss}
\end{equation}
where $\lambda(t)$ increases monotonically during training. Early training emphasizes semantic alignment, while later training shifts toward retrieval discriminability.

\newcommand{\gimp}[1]{\textcolor{green!60!black}{#1}} 
\newcommand{\deltas}[1]{{\scriptsize\gimp{#1}}}

\begin{table*}[t]
    \centering
    \setlength{\tabcolsep}{4pt} 
    \small 
\begin{adjustbox}{max width=\linewidth}
    
    \begin{tabular}{l l c c c c c c}
        \toprule
        \textbf{Model} & \textbf{Avg. (Ret.)} & \textbf{Time} $\downarrow$ & \textbf{FLOPs} $\downarrow$ & \textbf{Speedup} $\uparrow$ & \textbf{Exit Rate} $\uparrow$ & \textbf{Vis. Tok.} $\downarrow$ & \textbf{Txt. Tok.} $\downarrow$ \\
        \midrule

        Qwen2.5-VL-3B + CL & 68.9 (100\%) & 678.4s & 100\% & 1.00$\times$ & 0\% & 100\% & 100\% \\

        \textbf{Qwen2.5-VL-3B + SAS } & \textbf{68.0 (99\%)} & \textbf{471.3s} \deltas{$\downarrow$30.5\%} & \textbf{36.6\%} \deltas{$\downarrow$63.4\%} & \textbf{1.44$\times$} & \textbf{20\%} \deltas{$\uparrow$20\%} & \textbf{25\%} \deltas{$\downarrow$75\%} & \textbf{50\%} \deltas{$\downarrow$50\%} \\
        
        \midrule

        Qwen2.5-VL-7B + CL & 71.7 (100\%) & 1058.0s & 100\% & 1.00$\times$ & 0\% & 100\% & 100\% \\

        \textbf{Qwen2.5-VL-7B + SAS} & \textbf{70.9 (99\%)} & \textbf{644.2s} \deltas{$\downarrow$39.1\%} & \textbf{33.7\%} \deltas{$\downarrow$66.3\%} & \textbf{1.64$\times$}  & \textbf{26\%} \deltas{$\uparrow$26\%} & \textbf{25\%} \deltas{$\downarrow$75\%} & \textbf{50\%} \deltas{$\downarrow$50\%} \\
        
        \bottomrule
    \end{tabular}
\end{adjustbox}
    
    \caption{Efficiency analysis of Qwen2.5-VL-3B/7B with standard contrastive learning (CL) and their SAS-accelerated counterparts. 
Avg. (Ret.) reports the mean Hit@1 and the corresponding performance retention rate relative to the CL baseline. 
Exit Rate indicates the proportion of early-exited samples, while Vis./Txt. Tok. denote the ratios of retained visual and textual tokens after skimming.}

    \label{tab:efficiency}
\end{table*}

To preserve final-layer performance, we retain the standard contrastive objective at layer $L$. The overall Stage-I loss is
\begin{equation}
\mathcal{L}_{\text{StageI}}
=
\mathcal{L}_{con}^{(L)}
+
\gamma \mathcal{L}_{inter}(t),
\end{equation}
where $\gamma$ controls the strength of the intermediate supervision. After this stage, layer $l_d$ is explicitly trained as a usable retrieval representation, rather than remaining only a latent carrier of retrieval signals.

\paragraph{Stage II: Learning a Depth Sufficiency Assessor.}
Once the intermediate representation at layer $l_d$ has matured sufficiently, we train a depth assessor to determine whether it is sufficient for reliable retrieval. In this stage, the backbone is frozen, and only a lightweight decision module is optimized.

The assessor does not predict the retrieval result itself. Instead, it predicts whether the current representation has accumulated enough evidence for reliable matching, or whether deeper refinement is still needed. We formulate this as a binary classification problem over whether the sample should terminate at layer $l_d$.

The assessor uses three types of features. First, \emph{uncertainty scalars}, including the Top-1 similarity score and the Top-1/Top-2 similarity margin from intermediate retrieval results, capture ranking confidence. Second, \emph{modality indicators} distinguish text, image, and multimodal inputs, allowing the classifier to adapt to task-dependent distribution shifts. Third, \emph{semantic features} are obtained by projecting the intermediate \texttt{[EOS]} representation to encode the current semantic context. Full architectural and training details are provided in Appendix~\ref{app:classifier_details}.

\section{Experiment}

\begin{table*}[t]
\centering
\small
\setlength{\tabcolsep}{2.0pt}
\renewcommand{\arraystretch}{1.1}
\resizebox{\textwidth}{!}{%
\begin{tabular}{lccccccccccccc}
\toprule
\multirow{2}{*}{\textbf{Methods}}
& \multicolumn{4}{c}{$q^t \rightarrow c^i$}
& \multicolumn{2}{c}{$q^i \rightarrow c^t$}
& $q^i \rightarrow c^i$
& \multicolumn{2}{c}{$q^t \rightarrow (c^i, c^t)$}
& \multicolumn{2}{c}{$(q^i, q^t) \rightarrow c^i$}
& $(q^i, q^t) \rightarrow (c^i, c^t)$
& \multirow{2}{*}{Avg.} \\

\cmidrule(lr){2-5}
\cmidrule(lr){6-7}
\cmidrule(lr){8-8}
\cmidrule(lr){9-10}
\cmidrule(lr){11-12}
\cmidrule(lr){13-13}

& VisDial
& VN
& COCO
& Wiki-SS
& VN
& COCO
& NIGHTS
& WebQA
& EDIS
& CIRR
& FIQ
& OVEN
& \\
\midrule

UniIR~\citep{wei2024uniir}
& 42.2 & 74.3 & 68.5 & 12.2 & 76.8 & 72.1 & 66.2 & 89.6 & 79.2 & 51.3 & \underline{40.2} & 69.4 & 61.8 \\
GME~\citep{Zhang2024GMEIU}
& 48.1 & 74.7 & 68.1 & \textbf{73.9} & 78.3 & 63.1 & 67.0 & 88.8 & \underline{91.8} & 44.2 & 32.9 & 72.3 & 66.9 \\
VLM2Vec-2B~\citep{Jiang2024VLM2VecTV}
& 74.3 & 73.1 & 73.4 & 54.2 & 73.7 & 68.5 & 66.3 & 85.9 & 81.2 & 46.8 & 14.0 & 68.3 & 65.0 \\
VLM2Vec-7B~\citep{Jiang2024VLM2VecTV}
& 81.9 & \underline{80.5} & 77.2 & 62.3 & \underline{81.2} & 73.9 & 67.6 & 88.3 & 85.7 & 51.1 & 17.1 & 66.5 & 69.4 \\
VLM2Vec-V2.0~\citep{Meng2025VLM2VecV2AM}
& 82.7 & 74.5 & 75.3 & 66.9 & 78.2 & 71.4 & 68.6 & \underline{90.6} & 84.1 & 57.5 & 19.5 & 64.3 & 69.6 \\
LamRA-Qwen2-7B~\citep{Liu2024LamRALM}
& 61.3 & 70.4 & 72.2 & \underline{69.7} & \textbf{83.9} & 73.7 & 65.6 & 81.0 & 85.9 & 51.7 & \textbf{42.0} & \underline{82.0} & 70.0 \\
LamRA-Qwen2.5-7B~\citep{Liu2024LamRALM}
& 62.5 & 70.1 & 65.7 & 67.0 & 74.2 & 71.1 & 64.4 & 85.7 & 78.7 & 44.7 & 33.4 & \textbf{84.8} & 66.9 \\
\midrule
LLaVA-7B + CL
& 81.6 & \textbf{88.9} & 76.3 & 59.7 & 80.1 & \textbf{75.4} & \underline{68.7} & 75.9 & 83.1 & 58.3 & 17.4 & 50.8 & 68.0 \\
LLaVA-7B + SAS (ours)
& 81.6 & 75.6 & 76.3 & 57.3 & 80.0 & 73.5 & \textbf{69.7} & 87.2 & 75.2 & 56.2 & 17.4 & 54.5 & 67.1 \\
\midrule
Qwen2.5-VL-3B + CL
& 83.3 & 73.8 & 76.2 & 61.3 & 76.5 & 73.7 & 62.0 & 89.6 & 90.3 & \underline{59.3} & 19.9 & 61.0 & 68.9 \\
Qwen2.5-VL-3B + SAS (ours)
& \underline{83.9} & 71.3 & 76.5 & 55.1 & 76.2 & 72.2 & 68.2 & \textbf{90.7} & 85.8 & 54.8 & 20.5 & 60.6 & 68.0 \\
\midrule

Qwen2.5-VL-7B + CL
& \underline{83.9} & 77.4 & \textbf{78.4} & \underline{71.1} & 78.6 & \underline{73.9} & 68.0 & 90.5 & \textbf{93.0} & 54.1 & 25.0 & 66.8 & \textbf{71.7} \\
Qwen2.5-VL-7B + SAS (ours)
& \textbf{84.6} & 75.1 & \underline{77.8} & 64.5 & 78.8 & \underline{74.4} & 68.0 & 90.1 & 90.0 & \textbf{59.7} & 22.5 & 65.6 & \underline{70.9} \\
\bottomrule
\end{tabular}}
\caption{
Performance comparison among Qwen2.5-VL-3B/7B with contrastive learning (CL), our SAS-accelerated variants, and state-of-the-art retrieval methods.
All results are reported in Hit@1.}
\label{tab:mmeb_retrieval}
\end{table*}

\paragraph{Implementation Details.}
We instantiate SAS on top of Qwen2.5-VL-Instruct (3B and 7B)~\citep{Qwen2.5-VL}, and additionally evaluate its generality on the LLaVA-style training architecture~\citep{liu2024llava}. 
Stage I is conducted on 8 NVIDIA H100 GPUs with LoRA~\citep{hu2022lora} ($r=16$) for parameter-efficient fine-tuning. We train on the retrieval subset of MMEB~\citep{Jiang2024VLM2VecTV} for 5,000 steps with a learning rate of $5\times10^{-5}$, a global batch size of 1,024, and a contrastive temperature of $\tau=0.02$. 
Stage II is performed on a single NVIDIA A100 GPU, where we freeze the backbone and train the termination classifier for 1,000 steps using Focal Loss~\citep{lin2017focal} with $\alpha=0.8$ and $\gamma=3.0$, a learning rate of $5\times10^{-4}$, and a batch size of 512. 
 $\lambda(t)$ is initialized at 0 and annealed to 1.0 with a cosine schedule, gradually shifting the objective from distillation-based alignment to contrastive discrimination. We apply Vision Pooling at layer 1, Influence-Aware Token Filtering at layer $10$ ($l_s$), and place adaptive depth allocation at layer $12$ ($l_d$). For text tokens, we prune 50\% of the sequence while preserving a 16-token guardrail.

\paragraph{Evaluation.}
We evaluate SAS on the MMEB retrieval split, covering 12 benchmarks across five representative multimodal retrieval task types. Following standard practice, we report Hit@1 as the accuracy metric and measure end-to-end inference latency on a single NVIDIA A100 GPU.

\section{Main Results}
\label{sec:main_results}

We analyze the results from two perspectives: (1) efficiency relative to our reproduced dense models, and (2) retrieval performance compared with representative baselines.

\subsection{Efficiency Comparison}
\label{sec:efficiency_analysis}

To isolate the effect of SAS, we train dense 3B and 7B baselines under the same setting but without any adaptive acceleration, and use them as strict references for efficiency--accuracy comparison.

\paragraph{Efficiency--Accuracy Trade-off.}
As shown in Table~\ref{tab:efficiency}, SAS substantially reduces computation while preserving retrieval quality. Compared with the dense models, SAS retains nearly 99\% of the dense baseline’s average retrieval performance (e.g., 71.7\% $\rightarrow$ 70.9\% for 7B, and 68.9\% $\rightarrow$ 68.0\% for 3B), while reducing FLOPs by up to 66.3\%. This supports our claim that a considerable portion of token- and depth-level computation in dense MLLM retrieval is redundant and can be removed adaptively.

\paragraph{Revisiting the Parameter--Latency Trade-off.}
A notable finding is that the SAS-accelerated 7B model is even faster than the dense 3B model in end-to-end latency. In addition to outperforming the dense 3B model by about 2.0\% in retrieval accuracy, the accelerated 7B model also achieves lower inference time. We attribute this to stronger representation quality in the larger model, which enables more reliable identification of easy queries and therefore more frequent early termination at $l_d$. This result suggests that, under adaptive inference, a larger model can simultaneously maintain better accuracy and achieve lower latency than a smaller dense counterpart.

\subsection{Performance Comparison}
\label{sec:performance_comparison}

Table~\ref{tab:mmeb_retrieval} compares SAS with representative baselines, including dual-encoder retrievers and recent fine-tuned MLLM-based methods.

Despite aggressive token reduction and adaptive depth allocation, SAS remains highly competitive across diverse retrieval tasks. Our 7B model achieves an average Hit@1 of 70.9\%, remaining competitive with or outperforming strong recent MLLM-based retrieval baselines. This shows that adaptive sparse inference can significantly improve efficiency without compromising the representational strength of large retrieval models.

SAS is particularly effective on tasks requiring fine-grained semantic matching. On benchmarks such as VisDial and WebQA, it consistently outperforms prior methods. We attribute this to the complementary roles of Skim and Skip: Skim suppresses irrelevant information to sharpen semantic focus, while Skip preserves deeper reasoning for genuinely difficult queries.

\section{Discussion and Ablation}
\label{sec:discussion_ablation}

\subsection{Qualitative Analysis of Influence-Aware Token Scoring}
\label{sec:ablation_influence}

To better understand the behavior of the influence score $\psi_l(j)$ used in Skim, we visualize both its distributional behavior and its instance-level spatial selection pattern on the OVEN dataset, based on SAS-3B under 75\% visual-token reduction, collecting score statistics from 1,000 queries (112,554 visual tokens in total).

\begin{figure*}[t]
    \centering
    \begin{subfigure}{0.45\linewidth}
        \centering
        \includegraphics[width=\linewidth]{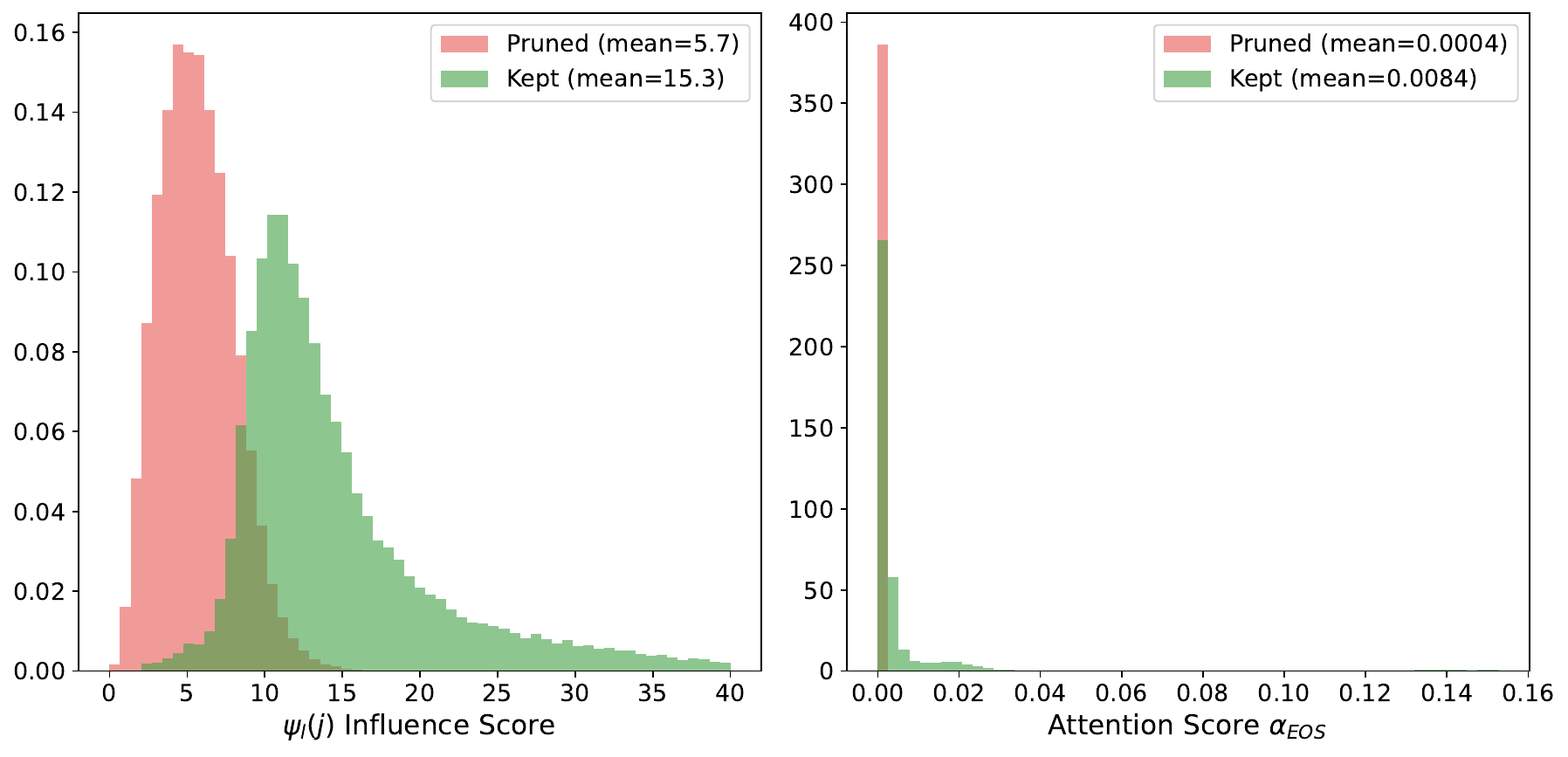}
        \caption{Score distribution.}
        \label{fig:score_dist}
    \end{subfigure}
    \hfill 
    \begin{subfigure}{0.51\linewidth}
        \centering
        \includegraphics[width=\linewidth]{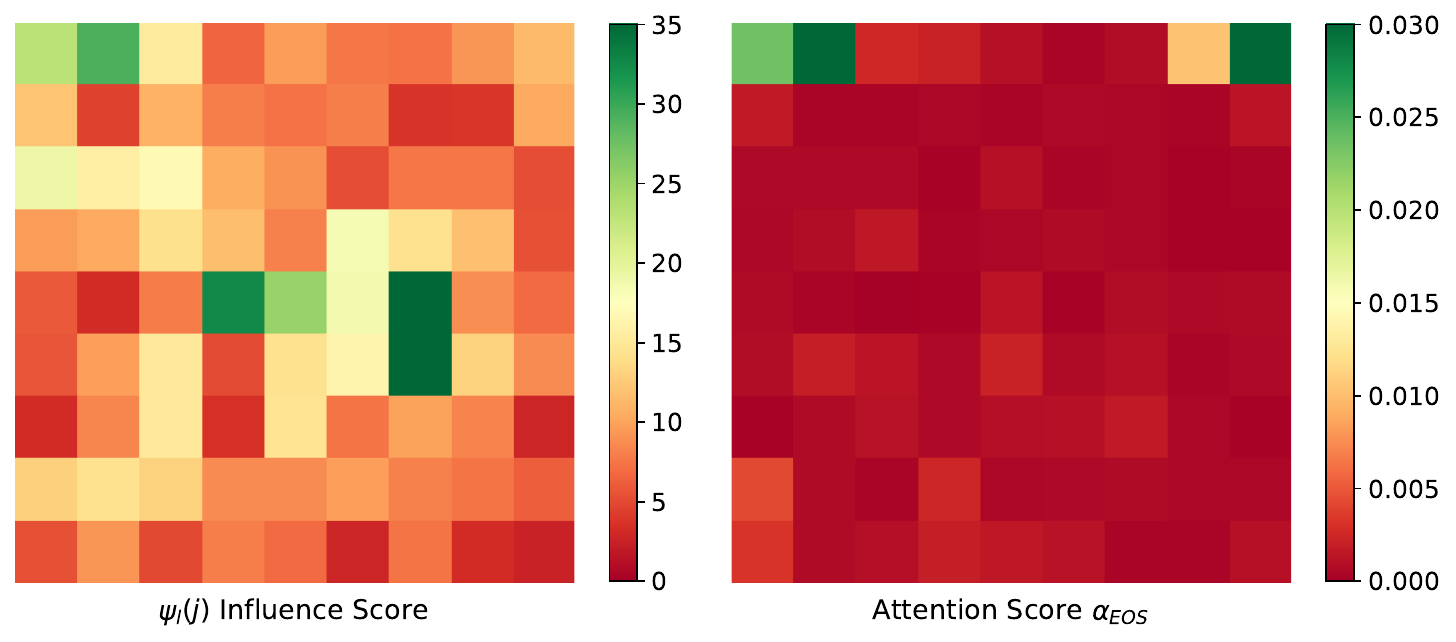}
        \caption{Spatial score map.}
        \label{fig:sample_scores}
    \end{subfigure}
    \caption{Qualitative comparison of influence-aware scoring vs.\ raw \textnormal{\texttt{[EOS]}} attention.}
    \label{fig:skim_vis}
\end{figure*}

As shown in Figure~\ref{fig:score_dist}, the proposed influence score yields a much clearer separation between kept and pruned tokens, with limited distributional overlap. In contrast, raw \texttt{[EOS]} attention collapses most tokens into a narrow range near zero, making the two groups barely distinguishable. This suggests that raw attention, as an instantaneous per-head signal, does not provide a stable pruning boundary. By aggregating perturbation effects across heads and accounting for downstream propagation, the influence score more faithfully captures each token's marginal contribution to the retrieval embedding.

Figure~\ref{fig:sample_scores} presents a representative spatial score map. The influence score highlights broader and more spatially coherent high-score regions, whereas raw attention responds only to a few isolated peaks. This indicates that retrieval-relevant evidence is often distributed across multiple tokens that jointly shape the \texttt{[EOS]} representation, rather than concentrated in a small number of extreme attention maxima---a pattern that generic attention-based pruning can easily miss.

We further ablate two design choices behind Skim. The selection layer $l_s$ determines when evidence filtering is applied: pruning too early (layers 6--8) removes tokens before cross-modal representations have sufficiently matured, while pruning too late (layer 13) leaves little room before the depth sufficiency check. Setting $l_s=10$ achieves the best trade-off; full results are provided in Appendix~\ref{app:pivot_layer_selection}. We also compare Skim against alternative pruning strategies in Appendix~\ref{app:prune_comparison}. Under 75\% visual and 50\% textual reduction, SAS retains 68.0\% performance, whereas vision-only methods (FastV and DART) drop to 39.3--60.6\%, and random text pruning falls to 58.3\%.

\vspace{-0.1cm}
\subsection{Exit Layer Selection}
\label{sec:ablation_exit_layer}

The exit layer $l_d$ determines where the depth sufficiency check is applied. A shallower exit saves more computation but may act on immature representations, whereas a deeper exit is more reliable but reduces the opportunity for early termination.

\begin{figure}[t]
    \centering
    \includegraphics[width=\linewidth]{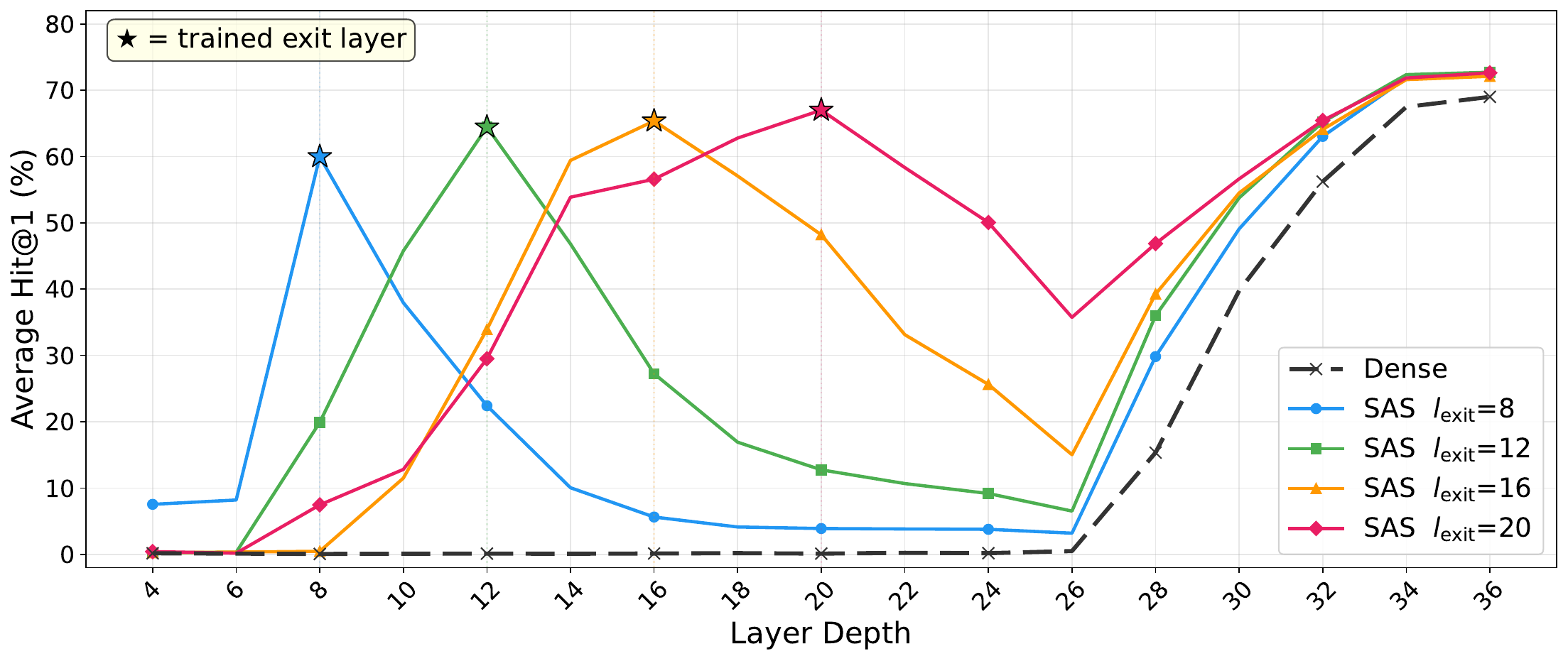}
    \caption{Performance under different $l_d$ choices.} 
    \vspace{-0.2cm}
    \label{fig:exit_layer_search}
\end{figure}

Figure~\ref{fig:exit_layer_search} compares per-layer retrieval performance for SAS variants trained with $l_d \in \{8,12,16,20\}$. Supervising a specific layer consistently boosts retrieval performance at that depth (marked by $\star$), confirming that Stage I effectively activates retrieval capability at the target layer. Among all choices, $l_d=12$ gives the best trade-off: the intermediate layer reaches 50.6\% average Hit@1, while the final layer achieves 69.4\%. Earlier exits yield weaker intermediate retrieval representations, whereas overly late exits diminish the practical benefit of adaptive termination. In particular, setting $l_d=13$ performs poorly, likely because there is too little depth remaining after the selection layer.

The full quantitative comparison is provided in Table~\ref{tab:exit_layer_ablation}. We further verify this effect through embedding-space visualization in Appendix~\ref{app:supervision_ablation}: after Stage I training, queries and their ground-truth candidates form much more coherent semantic clusters at $l_d$, whereas they are clearly separated before training. Additional comparisons of different supervision strategies are also included there.

\vspace{-0.1cm}
\subsection{Component Analysis}
\label{sec:ablation_components}

\begin{table}[t]
\centering
\small
\setlength{\tabcolsep}{6pt}
\begin{tabular}{cc|c|cc}
\toprule
\multicolumn{2}{c|}{\textbf{Components}} 
& \textbf{Performance} 
& \multicolumn{2}{c}{\textbf{Efficiency}} \\
\textbf{Skim} 
& \textbf{Skip} 
& \textbf{Avg.} $\uparrow$ 
& \textbf{FLOPs} $\downarrow$ 
& \textbf{Speedup} $\uparrow$ \\
\midrule
\ding{55} & \ding{55} & 68.9 & 100.0\% & 1.00$\times$ \\
\midrule
\ding{51} & \ding{55} & 68.0 & 42.2\%  & 1.26$\times$ \\
\ding{55} & \ding{51} & 68.1 & 87.9\%  & 1.17$\times$ \\
\ding{51} & \ding{51} & \textbf{68.0} & \textbf{36.6\%} & \textbf{1.44$\times$} \\
\bottomrule
\end{tabular}
\caption{Ablation Study of SAS Components.}
\label{tab:ablation_components}
\end{table}

Table~\ref{tab:ablation_components} reports a progressive ablation of the two hierarchical decisions in SAS. Evidence selection alone brings a 1.26$\times$ speedup with minimal performance loss (68.0\% vs.\ 68.9\%), indicating substantial token redundancy, while depth adaptation alone yields a 1.17$\times$ speedup, suggesting that many queries can terminate before the final layer. Combining the two leads to a 1.44$\times$ speedup at the same 68.0\% accuracy, showing that token selection and depth allocation address complementary sources of computation and work effectively together.

\section{Conclusion}
\label{sec:conclusion}
In this paper, we introduce \textbf{Skim and Skip (SAS)}, a hierarchical adaptive inference framework designed for universal multimodal retrieval. 
To overcome the computational bottlenecks inherent in standard dense retrieval, SAS reconfigures the inference paradigm by integrating two complementary strategies: 
\textit{Influence-Aware Token Filtering}, which effectively prunes redundant tokens to mitigate quadratic complexity, 
and \textit{Difficulty-Aware Early Exit}, which allows easy samples to terminate at intermediate layers.
Comparisons with strong dense MLLM backbones across multiple benchmarks demonstrate that SAS substantially improves inference efficiency while preserving competitive retrieval performance.

\clearpage

\section*{Limitations}
\label{sec:limitations}

Despite the encouraging results, we have not yet fully optimized the framework along two practical dimensions.

First, our current prototype uses standard dense attention kernels. Token skimming produces variable-length sequences. These irregular shapes are not always hardware-friendly. As a result, wall-clock gains can fall short of the theoretical FLOPs reduction. This is primarily an engineering gap. SAS is compatible with hardware-aware execution strategies. Examples include length-bucketed batching, fused indexing operators, and kernels that better support sparse or variable-length attention. We expect these improvements to narrow the theory–practice discrepancy.

Second, we evaluate SAS only on retrieval and embedding workloads. We have not validated it on multimodal generation tasks. Extending to VQA or captioning raises additional system constraints. In particular, KV-cache handling becomes more complex under varying token lengths. It also calls for more systematic choices of pruning and exit locations across backbones. Nevertheless, SAS introduces lightweight control without modifying the backbone. This design is data- and compute-efficient. We therefore expect that limited additional calibration, with small-scale task data or modest extra tuning compute, can enable robust configurations for generative settings.

\bibliography{custom}

\appendix
\newpage

\section{Detailed Related Work}
\subsection{Multimodal Retrieval}

Multimodal retrieval has evolved from dual-encoder models such as CLIP~\citep{Radford2021LearningTV} toward MLLM-based universal embedding frameworks. Early efforts such as E5-V~\citep{Jiang2024E5VUE}, VLM2Vec~\citep{Jiang2024VLM2VecTV}, and GME~\citep{Zhang2024GMEIU} demonstrated that MLLMs can be repurposed as strong zero-shot retrievers, while LamRA~\citep{Liu2024LamRALM} further improved alignment through retrieval-oriented tuning. More recent works continue to strengthen MLLM retrieval through universal embeddings~\citep{li2026magic,he2026plume}, flexible late interaction~\citep{xiao2025metaembed}, bottleneck representations~\citep{sun2026bottleneck}, and reasoning-enhanced retrieval features~\citep{cui2025reason,zhang2026reasoning}. For fine-grained scenarios, ColPali~\citep{Faysse2024ColPaliED} introduced multi-vector document representations derived directly from page images.

Beyond architectural design, another line of work improves retrieval representations through data scaling and training strategies. This includes expanding multimodal supervision via synthetic data~\citep{Zhou2024MegaPairsMD,chen2025mmE5,wu2026xcoderadvancingcompetitiveprogramming}, as well as improving contrastive learning with modality-aware curation~\citep{Kong2025ModalityCB}, smart hard-negative mining~\citep{Thirukovalluru2025BreakingTB}, and hardness-aware optimization~\citep{Lan2025LLaVELL}. Benchmark efforts such as MMEB-V3~\citep{huang2026mmeb} further highlight the remaining performance gaps of omni-modality embedding models. However, these approaches generally retain uniform dense inference at test time, applying the same amount of computation to all queries regardless of difficulty. In contrast, our work focuses on adaptive inference for multimodal retrieval by dynamically adjusting computation based on the query's information content and depth requirements.

\subsection{Token Compression and Adaptive Computation}

To reduce the computational cost of long multimodal sequences, prior work has extensively studied token compression, especially on the vision side. Attention-based methods such as FastV~\citep{Chen2024AnII}, SparseVLM~\citep{Zhang2024SparseVLMVT}, and ADSC~\citep{deniz2025adsc} rank tokens by importance scores, but often remain sensitive to local attention patterns. Similarity-based methods, including ToMe~\citep{Bolya2022TokenMY}, VisionZip~\citep{Yang2024VisionZipLI}, and DART~\citep{Wen2025StopLF}, merge or select representative tokens based on feature similarity, but are often weakly coupled to downstream retrieval objectives. More recent approaches incorporate stronger adaptivity through text conditioning, hierarchical reduction, or energy-aware scoring, as in CDPruner~\citep{Zhang2025BeyondAO}, GridPrune~\citep{duan2025gridprune}, HoloV~\citep{Zou2025DontJC}, HiDrop~\citep{wu2026hidrop}, and EAdaPrune~\citep{he2026energy}. Nevertheless, these methods mainly focus on reducing token count, typically under a predefined compression budget. Recent analysis also shows that visual token pruning can fail when relevant evidence shifts during inference~\citep{kim2026and}.

A complementary line of work studies adaptive computation through early exit or layer skipping. In language models, methods such as FastBERT~\citep{Liu2020FastBERTAS}, DeeBERT~\citep{Xin2020DeeBERTDE}, and ADEPT~\citep{Liu2020FastBERTAS, Xin2020DeeBERTDE, yoo2026adept, yang2024llm} dynamically reduce inference depth based on prediction confidence. Related ideas have also been explored in retrieval-augmented and multimodal settings~\citep{huang2024raee,yang2026dysl}. However, directly transferring these techniques to multimodal retrieval is non-trivial, since retrieval depends on the stability of continuous embeddings rather than discrete label predictions. Unlike prior work that treats token reduction and depth reduction separately, our method formulates efficient multimodal retrieval as a unified adaptive inference problem that jointly determines what information to keep and how deeply it should be processed.

\section{Theoretical Analysis of Influence-Aware Token Skimming}
\label{sec:appendix_theory}
In this section, we provide the complete derivation for the Influence-Aware Token Filtering score $\psi_l(j)$ (Eq.~(4) in the main text) and the bounding analysis for retrieval stability under token pruning.

\subsection{Derivation of Influence Score}
\label{app:influence_derivation}

\paragraph{Problem Formulation and Notation.}
We consider a Transformer-based MLLM backbone with $L$ layers. For a specific query encoding process, we focus on the self-attention mechanism at layer $l$ with $H$ heads.
Let $\mathbf{h}_l[i] \in \mathbb{R}^{D}$ denote the hidden state of token $i$ at layer $l$, and let $e$ denote the index of the [EOS] token used for pooling the final query embedding.
For a specific head $h$, the projection matrices are defined as $W_Q^{l,h}, W_K^{l,h}, W_V^{l,h} \in \mathbb{R}^{d \times D}$ and the output projection matrix as $W_O^{l,h} \in \mathbb{R}^{D \times d}$, where $d = D/H$.
Crucially, we incorporate the bias term for the value projection, denoted as $\mathbf{b}_V^{l,h} \in \mathbb{R}^d$.
Accordingly, the value vector for token $i$ is computed as:
\begin{equation}
    \mathbf{v}_{l,h,i} = W_V^{l,h} \mathbf{h}_l[i] + \mathbf{b}_V^{l,h},
\end{equation}
and the attention weight from the [EOS] token to token $i$ is denoted by $\alpha_{l,h}(e \to i)$.

\paragraph{Local Perturbation Analysis.}
Consider the self-attention output at the [EOS] position for a single head $h$ at layer $l$. The original output $\mathbf{o}_{l,h}(e)$ is a weighted sum of value vectors:
\begin{equation}
    \mathbf{o}_{l,h}(e) = \sum_{i} \alpha_{l,h}(e \to i) \mathbf{v}_{l,h,i}.
\end{equation}

When a token $j$ is removed (skimmed), the attention mechanism re-normalizes the weights of the remaining tokens. Assuming the pre-softmax logits remain unchanged for the surviving tokens, the new attention weight for any remaining token $i$ ($i \neq j$) becomes:
\begin{equation}
    \alpha'_{l,h}(e \to i) = \frac{\alpha_{l,h}(e \to i)}{1 - \alpha_{l,h}(e \to j)}.
\end{equation}

The perturbed output $\mathbf{o}'_{l,h}(e)$ after removing token $j$ is:
\begin{equation}
    \mathbf{o}'_{l,h}(e) = \sum_{i \neq j} \frac{\alpha_{l,h}(e \to i)}{1 - \alpha_{l,h}(e \to j)} \mathbf{v}_{l,h,i}.
\end{equation}

The exact perturbation vector $\Delta \mathbf{o}_{l,h}(e) = \mathbf{o}'_{l,h}(e) - \mathbf{o}_{l,h}(e)$ is derived as follows:
\begin{align}
\label{eq:delta_derivation_full}
\Delta \mathbf{o}_{l,h}(e)
&= \frac{\sum_{i \neq j} \alpha_{l,h}(e \to i) \mathbf{v}_{l,h,i}}{1 - \alpha_{l,h}(e \to j)} - \mathbf{o}_{l,h}(e) \notag \\
&= \frac{\mathbf{o}_{l,h}(e) - \alpha_{l,h}(e \to j) \mathbf{v}_{l,h,j}}{1 - \alpha_{l,h}(e \to j)} - \mathbf{o}_{l,h}(e) \notag \\
&= \frac{\alpha_{l,h}(e \to j)}{1 - \alpha_{l,h}(e \to j)} (\mathbf{o}_{l,h}(e) - \mathbf{v}_{l,h,j}).
\end{align}

This recovers Eq.~\eqref{eq:local_shift} presented in the main text.

\paragraph{Upper Bound on Influence Score.}
To efficiently select tokens for pruning, we derive a computable upper bound for the magnitude of the perturbation projected by $W_O^{l,h}$. We utilize the spectral norms of the weight matrices, denoted as $\sigma_V^{l,h} = \|W_V^{l,h}\|_2$ and $\sigma_O^{l,h} = \|W_O^{l,h}\|_2$.

First, we bound the term $\|\mathbf{o}_{l,h}(e) - \mathbf{v}_{l,h,j}\|_2$ using the triangle inequality:
\begin{equation}
    \|\mathbf{o}_{l,h}(e) - \mathbf{v}_{l,h,j}\|_2 \le \|\mathbf{o}_{l,h}(e)\|_2 + \|\mathbf{v}_{l,h,j}\|_2.
\end{equation}

We expand the norm of the value vector $\mathbf{v}_{l,h,j}$:
\begin{equation}
\begin{split}
    \|\mathbf{v}_{l,h,j}\|_2 &= \|W_V^{l,h} \mathbf{h}_l[j] + \mathbf{b}_V^{l,h}\|_2 \\
    &\le \sigma_V^{l,h} \|\mathbf{h}_l[j]\|_2 + \|\mathbf{b}_V^{l,h}\|_2.
\end{split}
\end{equation}

Next, we expand the norm of the original output $\mathbf{o}_{l,h}(e)$. Noting that $\sum_i \alpha_{l,h}(e \to i) = 1$, the bias term sums directly:
\begin{align}
    \|\mathbf{o}_{l,h}(e)\|_2
    &= \left\| \sum_{i} \alpha_{l,h}(e \to i) (W_V^{l,h} \mathbf{h}_l[i] + \mathbf{b}_V^{l,h}) \right\|_2 \notag \\
    &\le \sigma_V^{l,h} \sum_{i} \alpha_{l,h}(e \to i) \|\mathbf{h}_l[i]\|_2 + \|\mathbf{b}_V^{l,h}\|_2 \notag \\
    &= \sigma_V^{l,h} S_{l,h} + \|\mathbf{b}_V^{l,h}\|_2,
\end{align}
where we defined $S_{l,h} := \sum_{i} \alpha_{l,h}(e \to i) \|\mathbf{h}_l[i]\|_2$.

Substituting these bounds back into the perturbation equation:
\begin{align}
    &\|W_O^{l,h} \Delta \mathbf{o}_{l,h}(e)\|_2 \notag \\
    &\le \sigma_O^{l,h} \frac{\alpha_{l,h}(e \to j)}{1 - \alpha_{l,h}(e \to j)} \left( \|\mathbf{o}_{l,h}(e)\|_2 + \|\mathbf{v}_{l,h,j}\|_2 \right) \notag \\
    &\le \sigma_O^{l,h} \frac{\alpha_{l,h}(e \to j)}{1 - \alpha_{l,h}(e \to j)}
    \Bigg[
        \left(\sigma_V^{l,h} S_{l,h} + \|\mathbf{b}_V^{l,h}\|_2\right) \notag \\
    &\quad + \left(\sigma_V^{l,h} \|\mathbf{h}_l[j]\|_2 + \|\mathbf{b}_V^{l,h}\|_2\right)
    \Bigg] \notag \\
    &= \sigma_O^{l,h} \frac{\alpha_{l,h}(e \to j)}{1 - \alpha_{l,h}(e \to j)} \notag \\
    &\quad \cdot \Bigg(
        \sigma_V^{l,h} \Big( S_{l,h} + \|\mathbf{h}_l[j]\|_2 \Big) + 2\|\mathbf{b}_V^{l,h}\|_2
    \Bigg).
\end{align}

Summing this upper bound across all $H$ heads yields the final Influence Score $\psi_{l}(j)$, as used in the token filtering algorithm.

\subsection{Theoretical Interpretation: Connection to Integrated Gradients}

The derived influence score $\psi_{l}(j)$ aligns theoretically with feature attribution methods. Specifically, the importance of a token $h_j$ to the output function $F$ can be formalized using Integrated Gradients (IG):
\begin{equation}
    IG_j(F) = (h_j - 0) \times \int_{\alpha=0}^{1} \frac{\partial F(\alpha h_j)}{\partial h_j} d\alpha .
\end{equation}

In the context of the attention mechanism, assuming a single-step linear approximation ($\alpha=1$) and the ``Attention-as-Gradient'' hypothesis, the gradient flow is proportional to the attention weight weighted by the value projection. Consequently, the perturbation term in Eq.~\ref{eq:delta_derivation_full} can be viewed as:
\begin{equation}
    \|\Delta o_{l,h}(e)\| \approx \| \underbrace{\alpha_{l,h}(e \to j) \cdot v_{l,h,j}}_{\text{Attribution}} \| .
\end{equation}

Therefore, our filtering strategy $\min_j \psi_l(j)$ is closely related to pruning the token with the minimal first-order contribution to the layer's output.

\subsection{Global Stability Analysis via Neural ODEs}

While the derivation above guarantees minimal \textit{local} perturbation, we further demonstrate that this error does not propagate catastrophically. We view the residual connections in the MLLM backbone as an Euler discretization of a continuous Ordinary Differential Equation (ODE):
\begin{equation}
    h_{l+1} = h_l + f(h_l) \iff \frac{dh(t)}{dt} = f(h(t), t)
\end{equation}
Let $h(t)$ represent the trajectory of the full token sequence and $\tilde{h}(t)$ represent the trajectory after pruning a token at time $t_l$. The initial discrepancy introduced by pruning is bounded by our influence score: $\|\tilde{h}(t_l) - h(t_l)\| \le \epsilon = \min_j \psi_l(j)$.

Assuming the network dynamics function $f$ (Transformer block) is $K$-Lipschitz continuous (satisfied by bounded activation functions and LayerNorm), the divergence between the trajectories at the final time $T$ (Layer $L$) is governed by \textbf{Gronwall's Inequality}:
\begin{equation}
    \|\tilde{h}(T) - h(T)\| \le \|\tilde{h}(t_l) - h(t_l)\| \cdot e^{K(T - t_l)}.
\end{equation}

This inequality suggests that the error introduced by our skimming strategy remains bounded under the stated assumptions. By minimizing the initial perturbation $\|\tilde{h}(t_l) - h(t_l)\|$ via our Influence Score, the deviation in the final retrieval embedding can be kept under control, rather than growing chaotically.

\section{Implementation Details of Termination Classifier}
\label{app:classifier_details}
In this section, we provide the technical specifications for the Termination Classifier $g(\cdot)$ utilized in Stage II.

\subsection{Input Feature Construction}
The classifier aggregates information from three distinct sources. We specifically design these features to capture inference uncertainty, input modality, and semantic context:

\begin{itemize}[wide=0\parindent, itemsep=0em, topsep=0em]
    \item \textbf{Confidence \& Uncertainty Scalars ($\mathbf{x}_{stat} \in \mathbb{R}^{27}$):} 
    We extract 27 scalar statistics from the intermediate layer's output. Crucially, to explicitly inform the classifier about ranking stability, we include:
    \begin{enumerate}
        \item[(1)] \textbf{Top-1 Similarity Score:} The cosine similarity of the highest-ranked candidate, serving as a proxy for absolute confidence.
        \item[(2)] \textbf{Top-1/Top-2 Margin:} The difference between the similarity scores of the top-1 and top-2 candidates. A larger margin typically indicates a decisive prediction, while a narrow margin suggests potential ambiguity.
    \end{enumerate}
    Other metrics include attention entropy, token density, and maximum attention scores. To rectify the highly skewed, long-tail distribution of these values (e.g., entropy varies significantly across layers), we apply a signed \texttt{Log1p} transformation followed by Layer Normalization:
    \begin{equation}
    \begin{aligned}
        \hat{\mathbf{x}}_{stat} = \text{LayerNorm} ( & \text{sign}(\mathbf{x}_{stat}) \\
        & \cdot \ln(1 + |\mathbf{x}_{stat}|) ).
    \end{aligned}
    \end{equation}

    \item \textbf{Modality Embedding ($\mathbf{x}_{mod} \in \mathbb{R}^{4}$):} 
    A learnable embedding layer maps the discrete modality index (representing text-only, image-only, or multimodal inputs) to a 4-dimensional continuous vector. This allows the classifier to adapt its decision boundary biases according to the specific input data type.

    \item \textbf{Semantic Projection ($\mathbf{x}_{sem} \in \mathbb{R}^{64}$):} 
    To incorporate semantic context without incurring high computational costs, the high-dimensional backbone embedding $\mathbf{e}_{l_{exit}}$ ($D=3584$ for Qwen2.5-VL) is projected to a lower dimension via a non-linear head:
    \begin{equation}
        \mathbf{x}_{sem} = \text{ReLU}(\text{LayerNorm}(\mathbf{W}_{p} \mathbf{e}_{l_{exit}} + \mathbf{b}_p)),
    \end{equation}
    where the projection dimension is set to 64.
\end{itemize}

\subsection{Network Architecture}
The processed features are concatenated into a single vector $\mathbf{h}_{in} \in \mathbb{R}^{95}$ and fed into the classifier. The architecture is a Multi-Layer Perceptron (MLP) defined as:
\begin{equation}
    p_{exit} = \text{Sigmoid}(\text{MLP}_{\text{LN}}(\mathbf{h}_{in})).
\end{equation}
Specifically, $\text{MLP}_{\text{LN}}$ consists of two hidden layers with dimensions 128 and 64, respectively. We consistently employ Layer Normalization instead of Batch Normalization after each linear transformation, followed by ReLU activation and Dropout ($p=0.2$). This design ensures that normalization statistics are computed independently for each instance, enhancing robustness against batch-size variations and distribution shifts during inference.

\subsection{Training Objective}
Since samples that are already sufficient at intermediate layers and those requiring deeper computation are typically imbalanced, we optimize the classifier using Focal Loss~\citep{lin2017focal}:
\begin{equation}
\mathcal{L}_{focal} = -\alpha(1 - p_t)^\gamma \log(p_t),
\end{equation}
where $p_t$ represents the estimated probability for the ground-truth class. This loss reduces the dominance of easy-to-classify samples, focusing training on boundary cases where the sufficiency decision is most ambiguous. We set $\alpha=0.8$ and $\gamma=3.0$ in all experiments.

\section{Hyperparameter Sensitivity: Selection Layer Analysis}
\label{app:pivot_layer_selection}
\subsection{Selection of the Selection Layer ($l_s$)}

A key design choice in SAS is the position of the selection layer $l_s$, where Influence-Aware Token Filtering is applied. Choosing $l_s$ requires balancing computational savings against representational maturity:
\begin{itemize}[wide=0pt, itemsep=0em, topsep=0em]
    \item \textbf{Too early ($l_s < 8$):} Multimodal representations may not yet be sufficiently mature. Pruning at this stage risks removing tokens that appear locally redundant but are still important for deeper cross-modal aggregation.
    \item \textbf{Too late ($l_s > 12$):} Although safer, delayed pruning preserves redundant tokens for too many layers and weakens the efficiency gains.
\end{itemize}

To identify an appropriate $l_s$, we conduct an ablation study on Qwen2.5-VL-3B trained for 2,000 steps on MMEB. We vary the selection layer over $l_s \in \{6, 8, 10, 11, 12, 13\}$ and measure retrieval performance (Hit@1) at both the intermediate layer ($L_{12}$) and the final layer ($L_{36}$).

\begin{table}[h]
    \centering
    \small
    \renewcommand{\arraystretch}{1.2}
    \setlength{\tabcolsep}{12pt}
    \begin{tabular}{ccc}
    \toprule
    \multirow{2}{*}{\textbf{Selection ($l_s$)}} & \multicolumn{2}{c}{\textbf{Retrieval Performance (Hit@1)}} \\
    \cmidrule(lr){2-3}
     & \textbf{Inter. ($L_{12}$)} & \textbf{Final ($L_{36}$)} \\
    \midrule
    6  & 39.5\% & 66.3\% \\
    8  & 39.3\% & 66.4\% \\
    \rowcolor{gray!15}
    \textbf{10} & 44.4\% & \textbf{69.2\%} \\
    11 & 44.4\% & 68.1\% \\
    12 & 45.4\% & 68.7\% \\
    13 & 5.3\%  & 68.2\% \\
    \bottomrule
    \end{tabular}
    \caption{\textbf{Impact of selection layer choice.} Retrieval performance at the intermediate and final layers under different $l_s$ settings. $l_s=10$ achieves the best balance.}
    \label{tab:pivot_ablation}
\end{table}

\paragraph{Analysis.}
As shown in Table~\ref{tab:pivot_ablation}, applying pruning too early (layers 6 or 8) noticeably degrades final performance (around 66.3\%--66.4\% versus 69.2\%), confirming that early representations are still too unstable for aggressive reduction.

Notably, \textbf{$l_s=10$} gives the best overall result, outperforming even later choices such as $l_s=12$. We conjecture that selecting tokens at layer 10 leaves a useful buffer before the first depth sufficiency check at $l_d=12$, allowing the remaining tokens to re-distribute attention and consolidate cross-modal semantics after pruning. By contrast, when filtering is applied too close to the first depth decision, the model has insufficient room to recover from the perturbation introduced by token removal. We therefore set $l_s=10$ in all main experiments.

\section{Token Pruning Strategy Comparison}
\label{app:prune_comparison}

We compare SAS with existing token pruning strategies applied to the DenseRet backbone, including vision-centric methods (VisionZip, FastV, DART) and a text-centric baseline (Random Text Pruning).

\begin{table}[t]
\centering
\small
\setlength{\tabcolsep}{3.5pt}
\renewcommand{\arraystretch}{1.05}
\resizebox{\columnwidth}{!}{
\begin{tabular}{lcccc}
\toprule
\textbf{Method} 
& \textbf{Avg.} $\uparrow$ 
& \textbf{Vis. R.} $\uparrow$ 
& \textbf{Text R.} $\uparrow$ 
& \textbf{FLOPs} $\downarrow$ \\
\midrule
DenseRet-3B           
& 68.9 & 0\% & 0\% & 100\% \\
\midrule
\multicolumn{5}{l}{\textit{Vision Pruning Baselines}} \\
\quad + VisionZip
& 39.3 & 75\% & 0\% & 48.5\% \\
\quad + FastV
& 60.6 & 75\% & 0\% & 51.3\% \\
\quad + DART        
& 59.0 & 75\% & 0\% & 51.3\% \\
\midrule
\multicolumn{5}{l}{\textit{Text Pruning Baseline}} \\
\quad + Random
& 58.3 & 0\% & 50\% & 76.9\% \\
\midrule
\multicolumn{5}{l}{\textit{Ours}} \\
\quad + SAS                 
& \textbf{68.0} & \textbf{75\%} & \textbf{50\%} & \textbf{36.6\%} \\
\bottomrule
\end{tabular}
}
\caption{
Comparison with vision and text token pruning strategies.
Vis. R. and Text R. denote the reduction rates of visual and textual tokens, respectively.}
\label{tab:token_pruning_comparison}
\end{table}

As shown in Table~\ref{tab:token_pruning_comparison}, vision-optimized methods cause severe degradation (e.g., VisionZip: 68.9\% $\to$ 39.3\%), as they prioritize local visual salience without reference to the retrieval objective. Random text pruning (58.3\%) indiscriminately removes tokens critical for semantic alignment. SAS sustains 68.0\% performance at substantially higher reduction ratios (75\% visual, 50\% textual) because its influence metric specifically preserves tokens with high marginal contribution to the \texttt{[EOS]} retrieval embedding.

\section{Exit Layer Ablation}
Table 8 reports the average Hit@1 at the selected decision layer and at the final layer under different choices of $l_d$. Although $l_d=10$ and $l_d=12$ achieve comparable final-layer performance, we use $l_d=12$ in the main experiments because it provides a more stable decision point after token selection and better practical early-exit behavior.
\label{app:exit_layer_ablation}
\begin{table}[t]
\centering
\small
\setlength{\tabcolsep}{7pt}
\begin{tabular}{ccc}
\toprule
\textbf{$l_d$} & \textbf{Hit@1 at $l_d$} & \textbf{Hit@1 at $L_{36}$} \\
\midrule
8  & 47.7 & 66.8 \\
10 & 50.9 & 69.4 \\
11 & 50.5 & 68.2 \\
\rowcolor{gray!15}\textbf{12} & 50.6 & \textbf{69.4} \\
13 & 18.8 & 67.7 \\
\bottomrule
\end{tabular}
\caption{\textbf{Impact of decision layer choice.} Average Hit@1 at the selected decision layer $l_d$ and at the final layer. We choose $l_d{=}12$ in the main experiments.}
\label{tab:exit_layer_ablation}
\end{table}

\section{Detailed Ablation on Supervision Strategies}
\label{app:supervision_ablation}
\begin{table}[h]
    \centering
    \small 
    \setlength{\tabcolsep}{5pt} 
    
    \caption{\textbf{Impact of Intermediate Supervision Strategies.} Comparison of Hit@1 accuracy at the intermediate ($L_{12}$) and final ($L_{last}$) layers.}
    \label{tab:app_loss_ablation}
    
    \begin{tabular}{l c c}
        \toprule
        \textbf{Training Strategy} & \textbf{Exit ($L_{12}$)} & \textbf{Final ($L_{last}$)} \\
        \midrule
        Baseline (No Loss) & 0.01 & 68.9 \\
        $+\mathcal{L}_{con}$ Only & 49.4 & 69.0 \\
        $+\mathcal{L}_{distill}$ Only & \textbf{57.4} & 69.2 \\
        \midrule
        \rowcolor{gray!10} \textbf{Hybrid (Ours)} & 57.2 & \textbf{69.5} \\
        \bottomrule
    \end{tabular}
\end{table}

This section provides the detailed numerical comparison of different intermediate supervision strategies. Table~\ref{tab:app_loss_ablation} reports the retrieval accuracy (Hit@1) at both the intermediate exit layer ($L_{12}$) and the final layer ($L_{last}$).
Further analysis reveals the limitations of single objectives. Applying solely $\mathcal{L}_{con}$ leads to under-fitting (49.4\%) due to optimization difficulties at shallow layers, while solely $\mathcal{L}_{distill}$ over-regularizes the network, suppressing the final layer's potential ($L_{last}$ drops to 69.2\%). In contrast, our Hybrid Strategy resolves this dilemma, securing high utility for early exiting (57.2\%) while boosting the final layer's peak performance to \textbf{69.5\%}.

\section{More Visualization Results}
\label{app:more_tsne}
\begin{figure}[htbp]
  \centering

  \begin{subfigure}{\columnwidth}
    \centering
    \includegraphics[width=0.49\columnwidth]{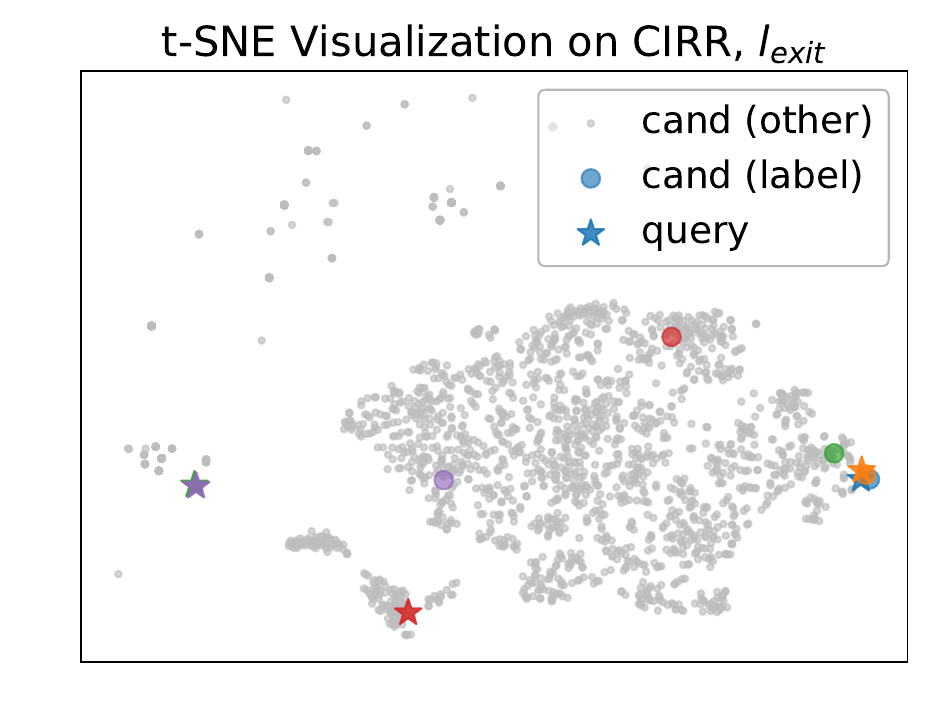}
    \hfill
    \includegraphics[width=0.49\columnwidth]{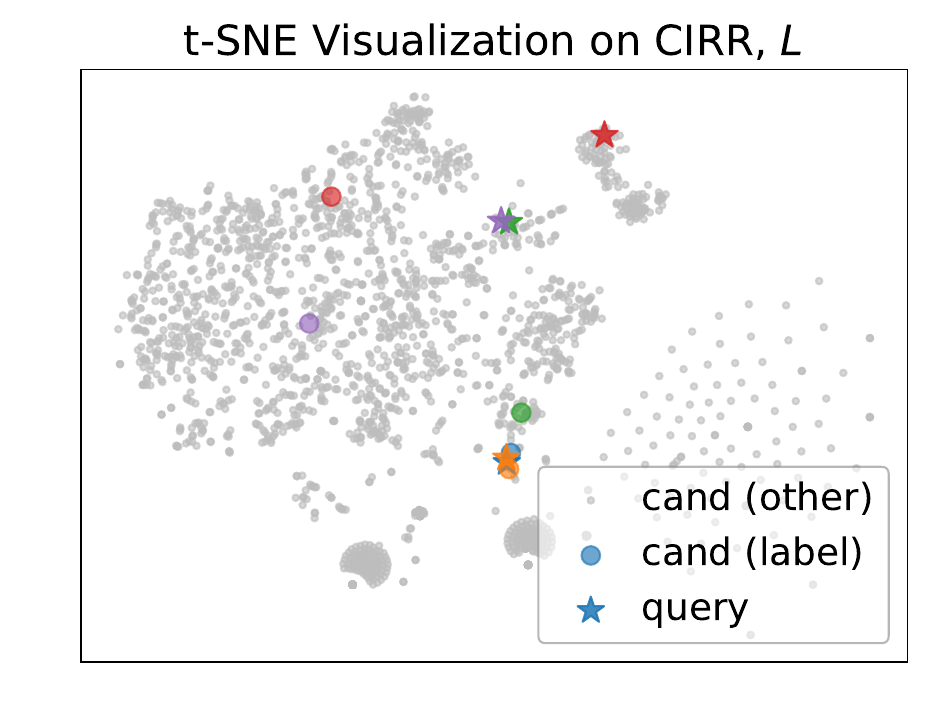}
    \caption{Before training}
  \end{subfigure}

  \vspace{0.5em}

  \begin{subfigure}{\columnwidth}
    \centering
    \includegraphics[width=0.49\columnwidth]{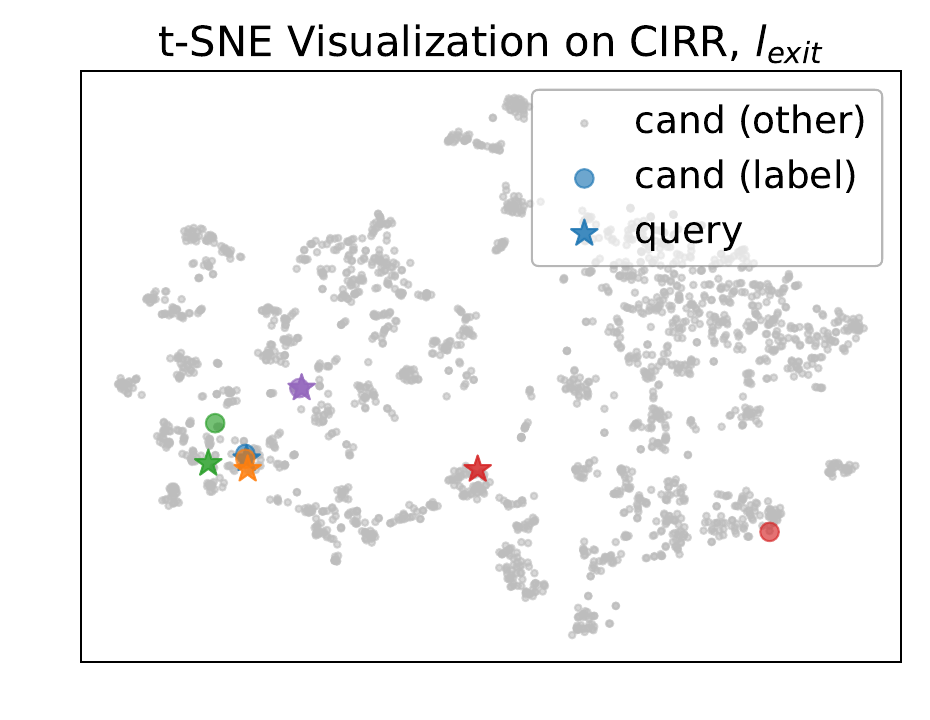}
    \hfill
    \includegraphics[width=0.49\columnwidth]{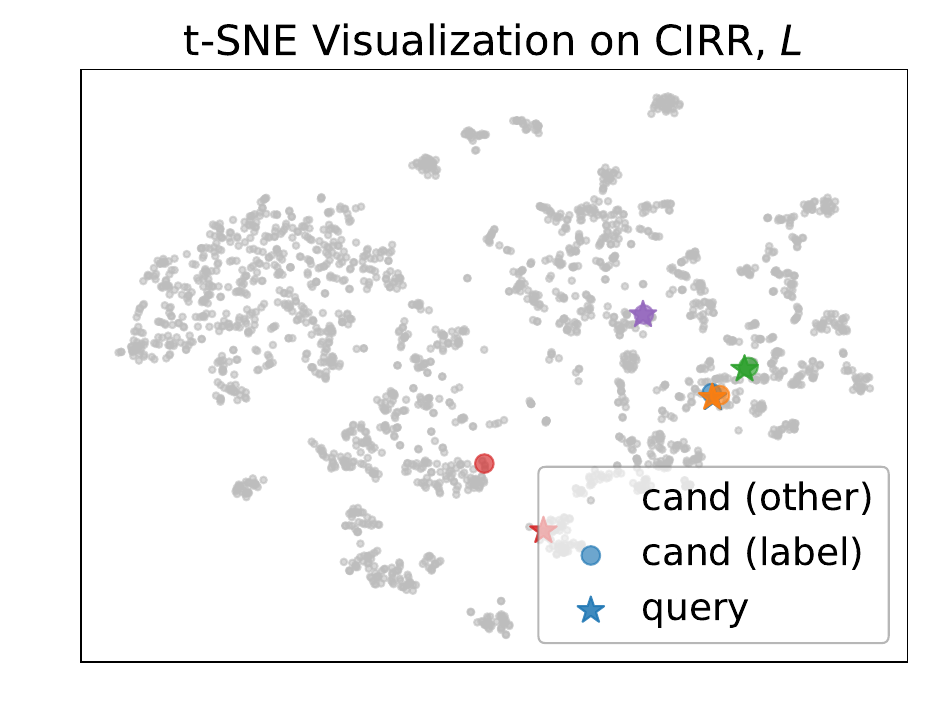}
    \caption{After training}
  \end{subfigure}

  \caption{t-SNE visualizations on CIRR}
\end{figure}

\begin{figure}[htbp]
  \centering

  \begin{subfigure}{\columnwidth}
    \centering
    \includegraphics[width=0.49\columnwidth]{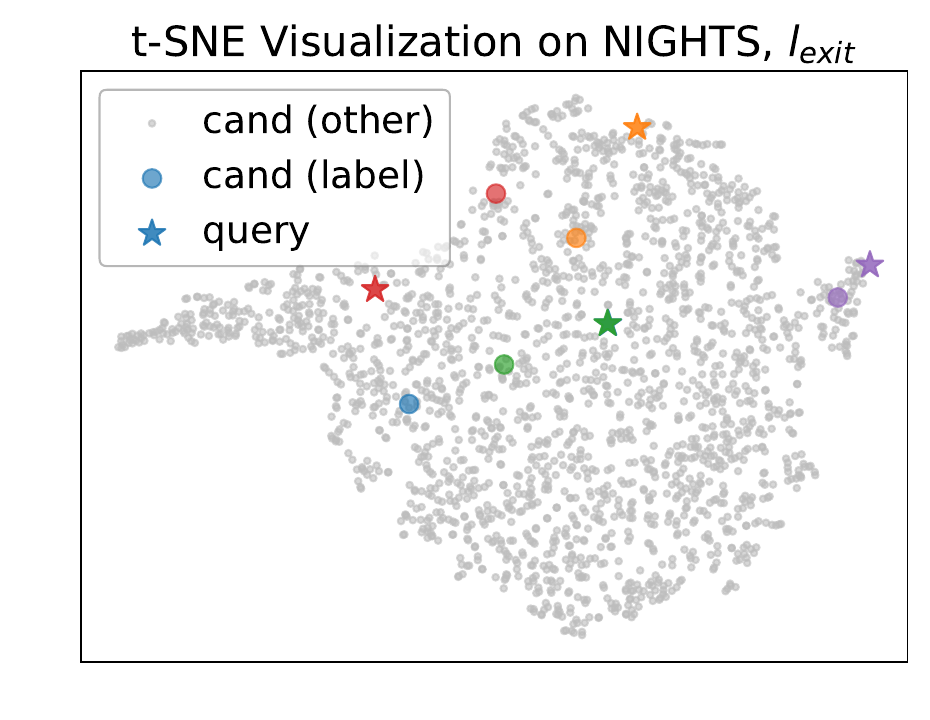}
    \hfill
    \includegraphics[width=0.49\columnwidth]{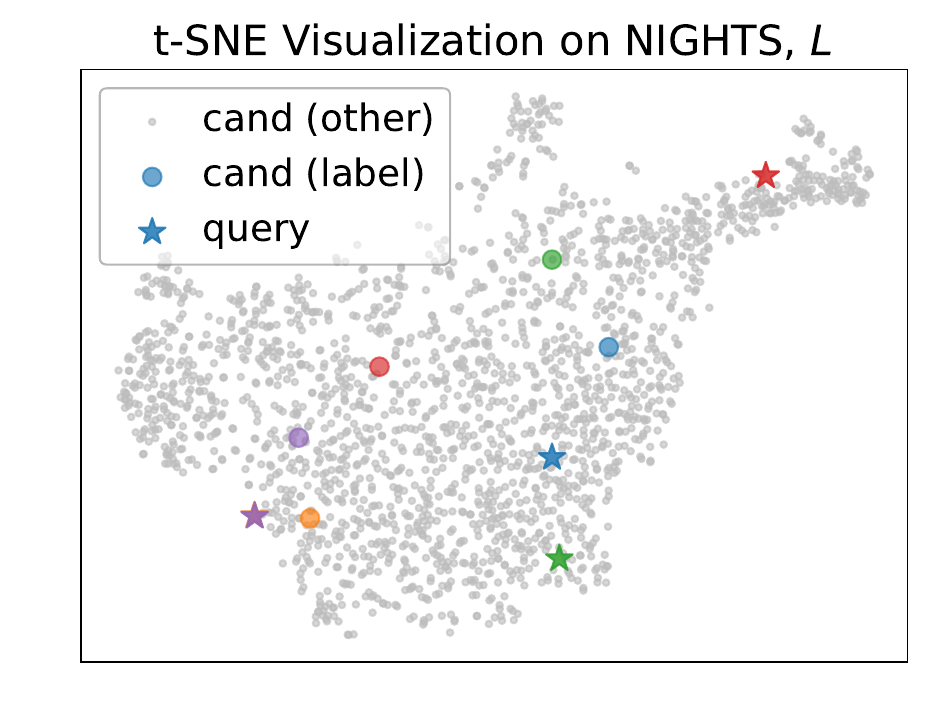}
    \caption{Before training}
  \end{subfigure}

  \vspace{0.5em}

  \begin{subfigure}{\columnwidth}
    \centering
    \includegraphics[width=0.49\columnwidth]{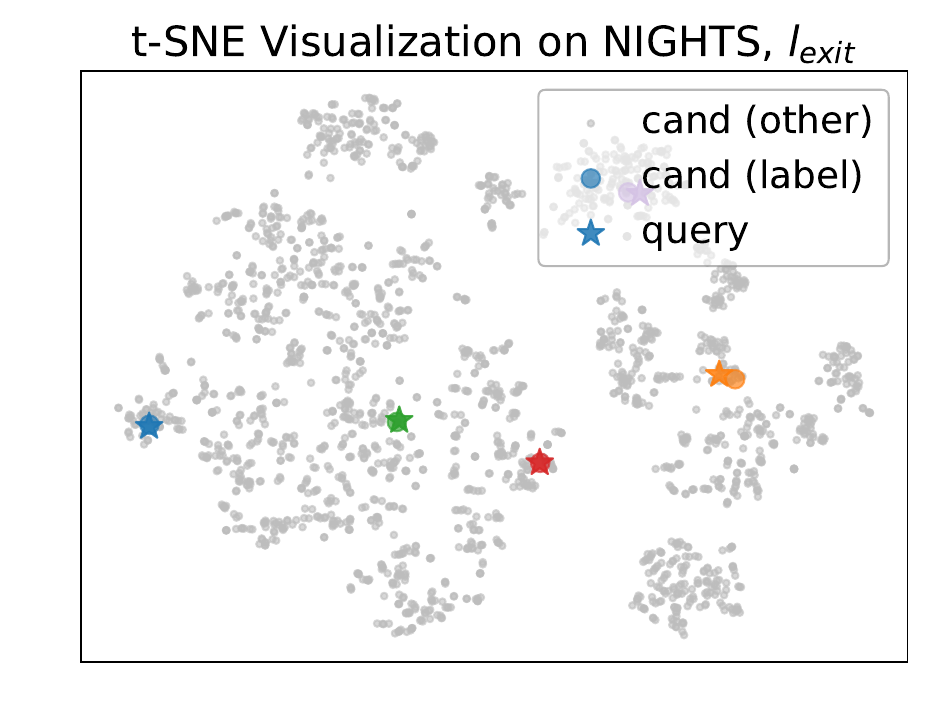}
    \hfill
    \includegraphics[width=0.49\columnwidth]{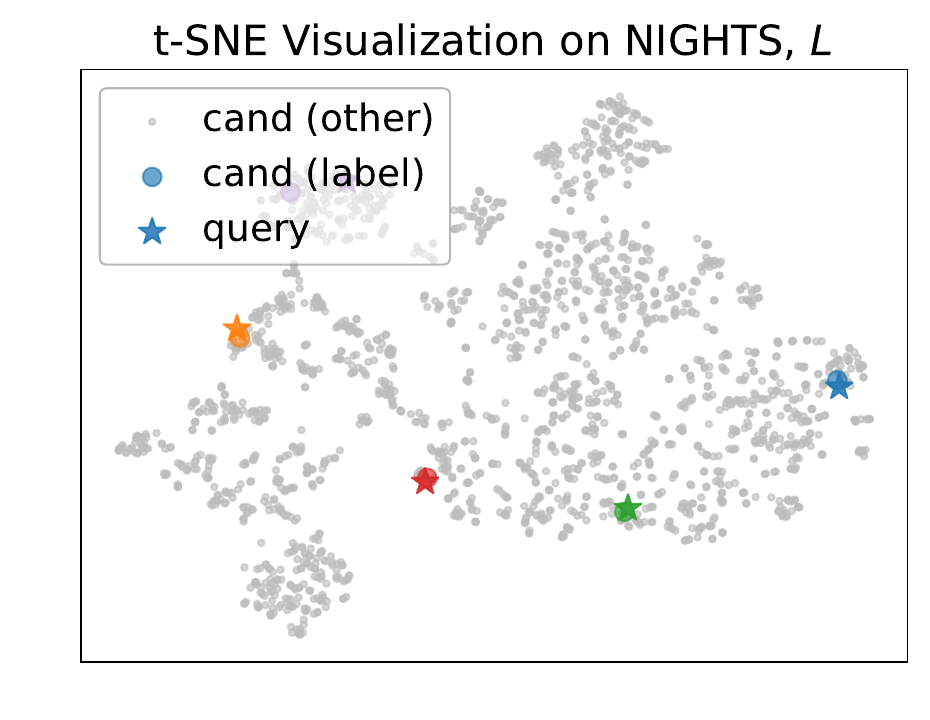}
    \caption{After training}
  \end{subfigure}

  \caption{t-SNE visualizations on NIGHTS}
\end{figure}

\begin{figure}[htbp]
  \centering

  \begin{subfigure}{\columnwidth}
    \centering
    \includegraphics[width=0.49\columnwidth]{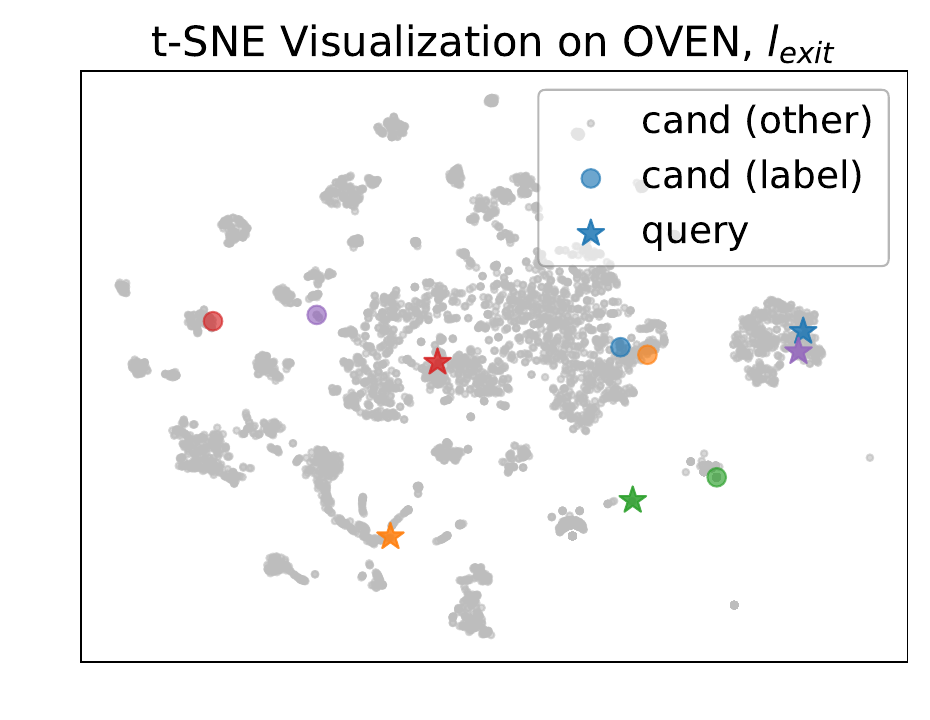}
    \hfill
    \includegraphics[width=0.49\columnwidth]{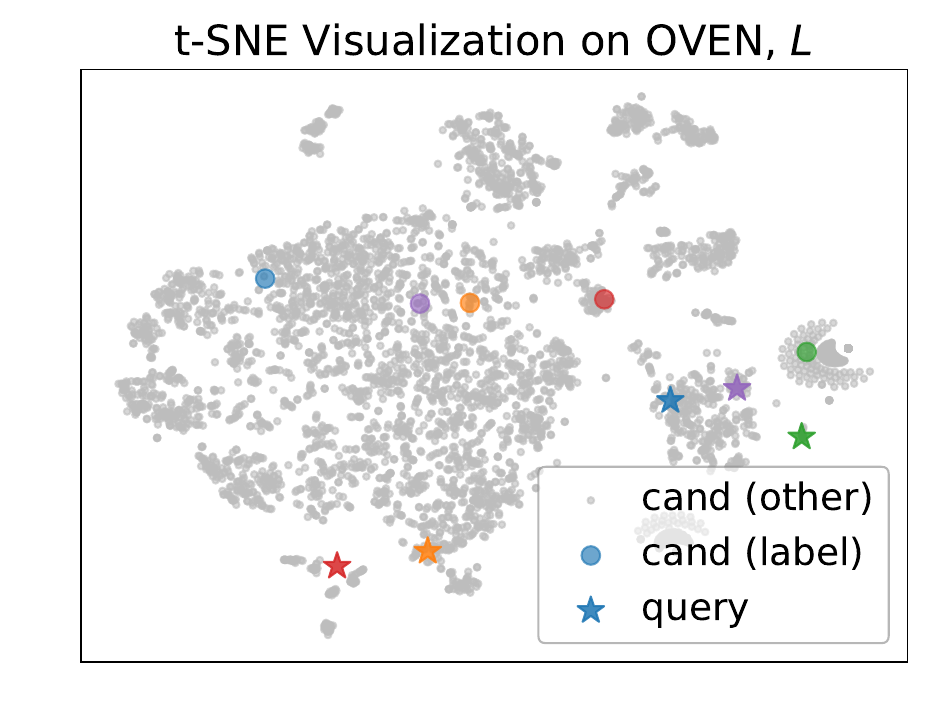}
    \caption{Before training}
  \end{subfigure}

  \vspace{0.5em}

  \begin{subfigure}{\columnwidth}
    \centering
    \includegraphics[width=0.49\columnwidth]{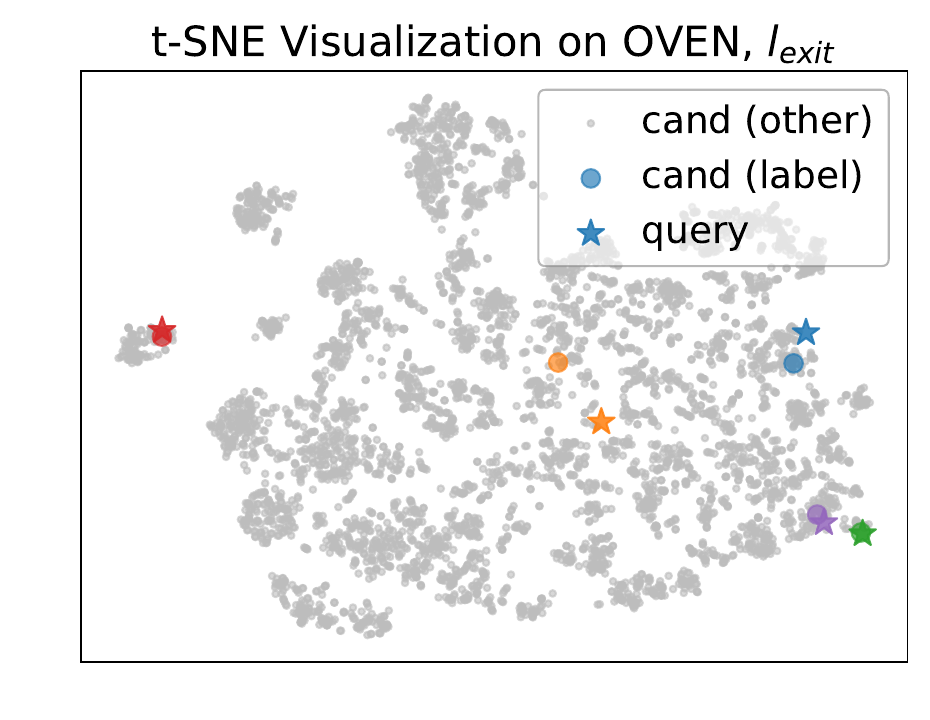}
    \hfill
    \includegraphics[width=0.49\columnwidth]{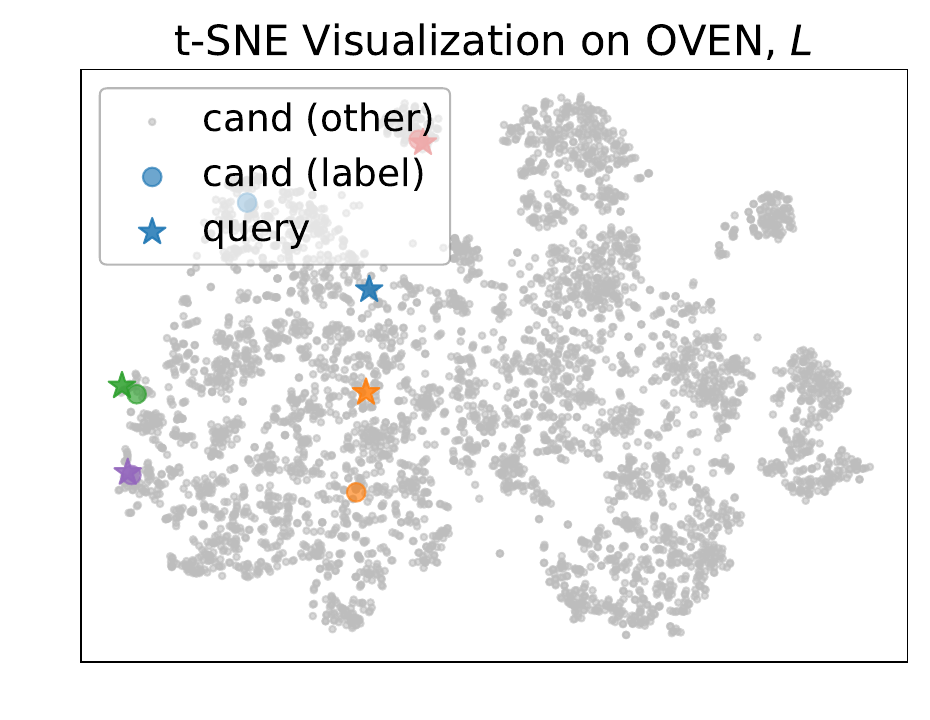}
    \caption{After training}
  \end{subfigure}

  \caption{t-SNE visualizations on OVEN}
\end{figure}

\begin{figure}[htbp]
  \centering

  \begin{subfigure}{\columnwidth}
    \centering
    \includegraphics[width=0.49\columnwidth]{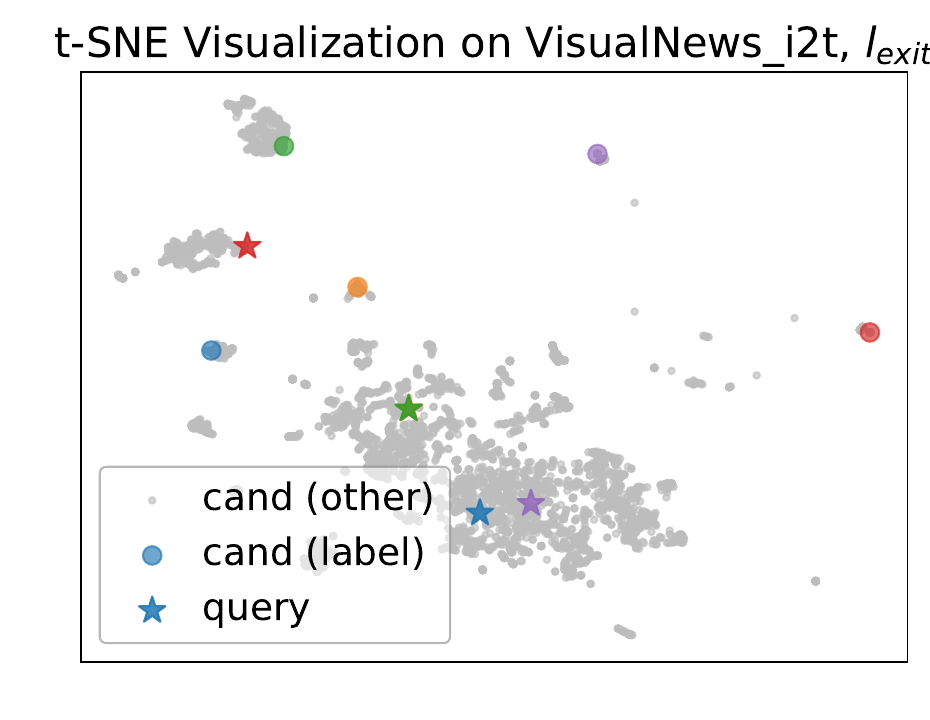}
    \hfill
    \includegraphics[width=0.49\columnwidth]{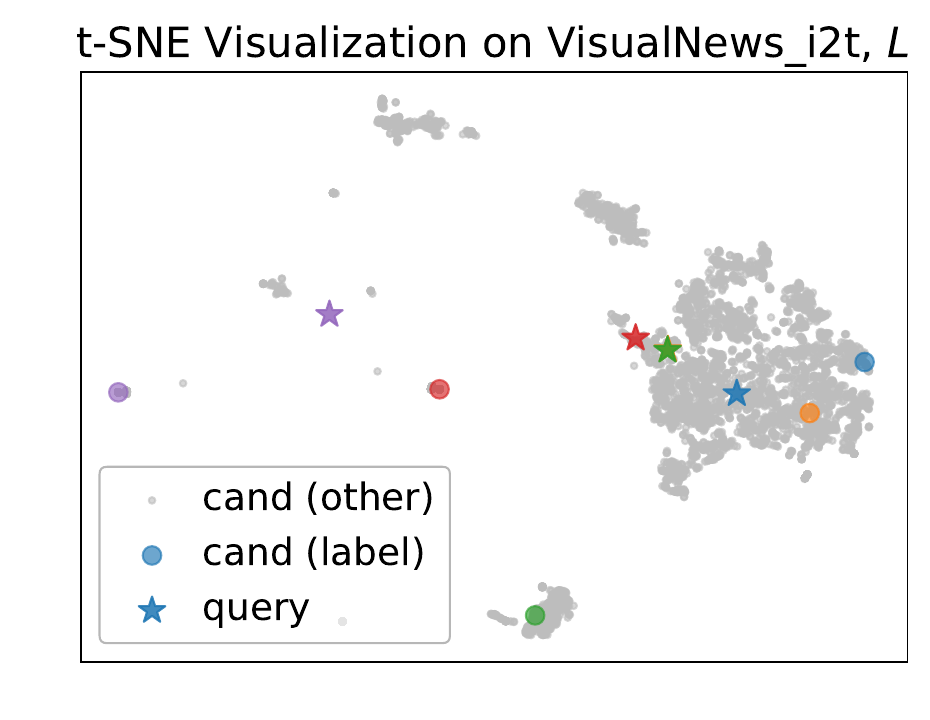}
    \caption{Before training}
  \end{subfigure}

  \vspace{0.5em}

  \begin{subfigure}{\columnwidth}
    \centering
    \includegraphics[width=0.49\columnwidth]{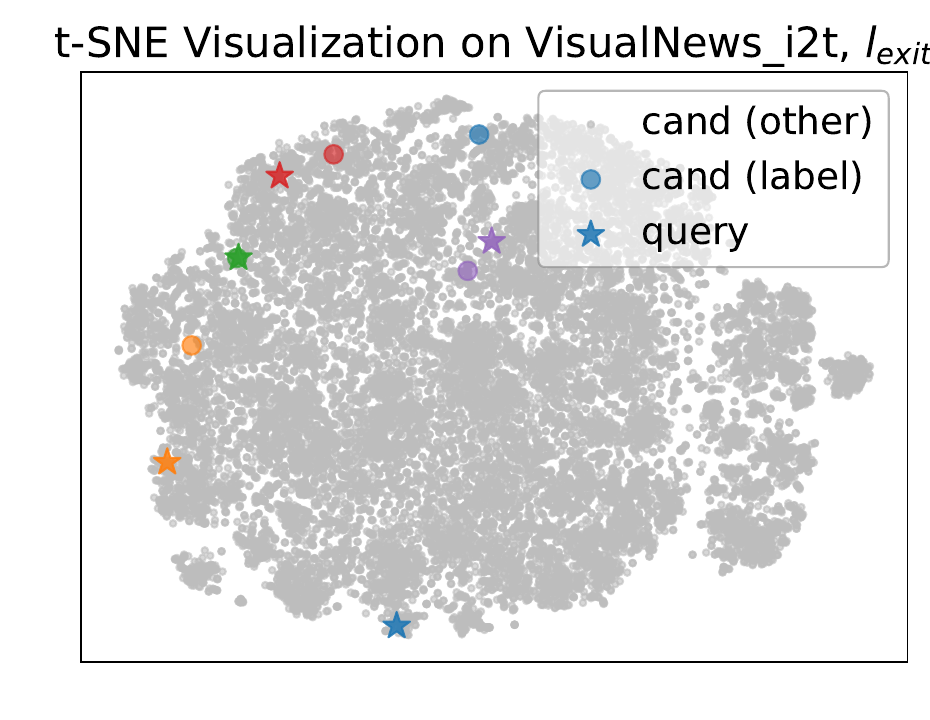}
    \hfill
    \includegraphics[width=0.49\columnwidth]{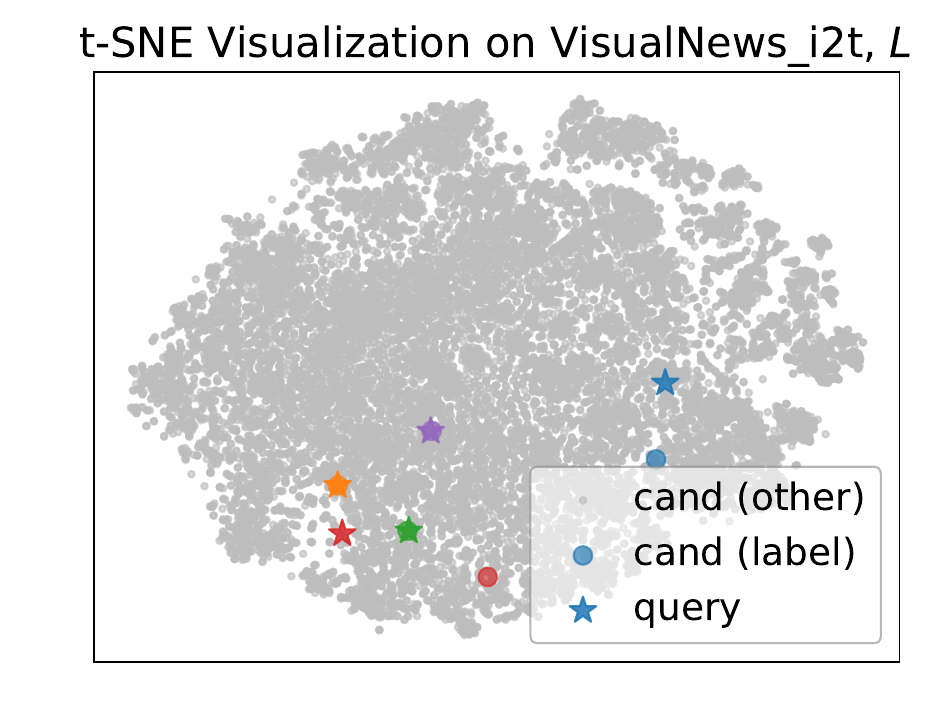}
    \caption{After training}
  \end{subfigure}

  \caption{t-SNE visualizations on VisualNews\_i2t}
\end{figure}

\begin{figure}[htbp]
  \centering

  \begin{subfigure}{\columnwidth}
    \centering
    \includegraphics[width=0.49\columnwidth]{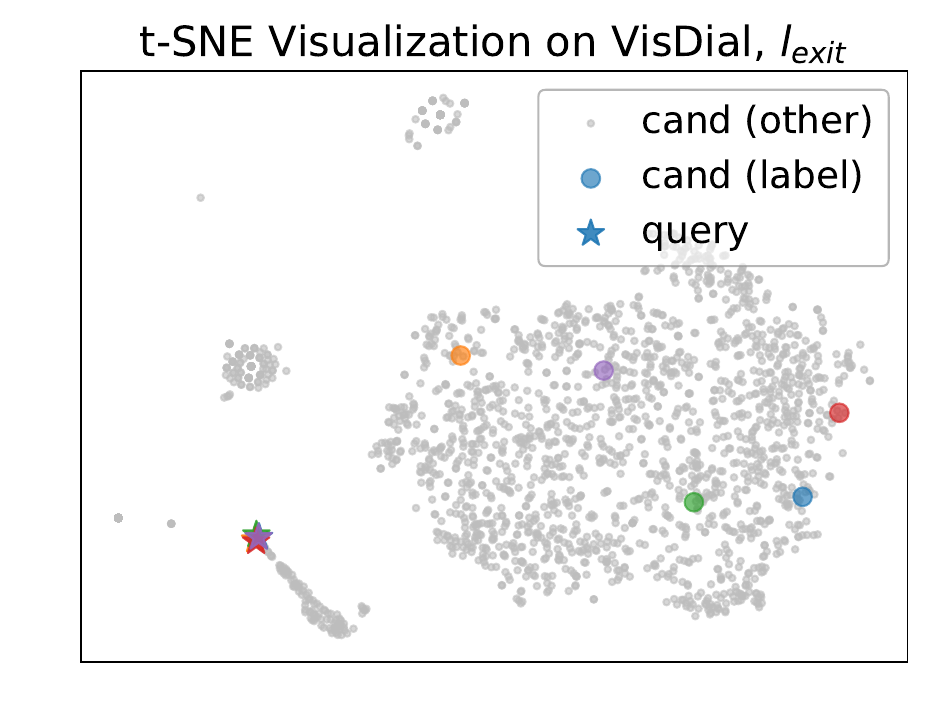}
    \hfill
    \includegraphics[width=0.49\columnwidth]{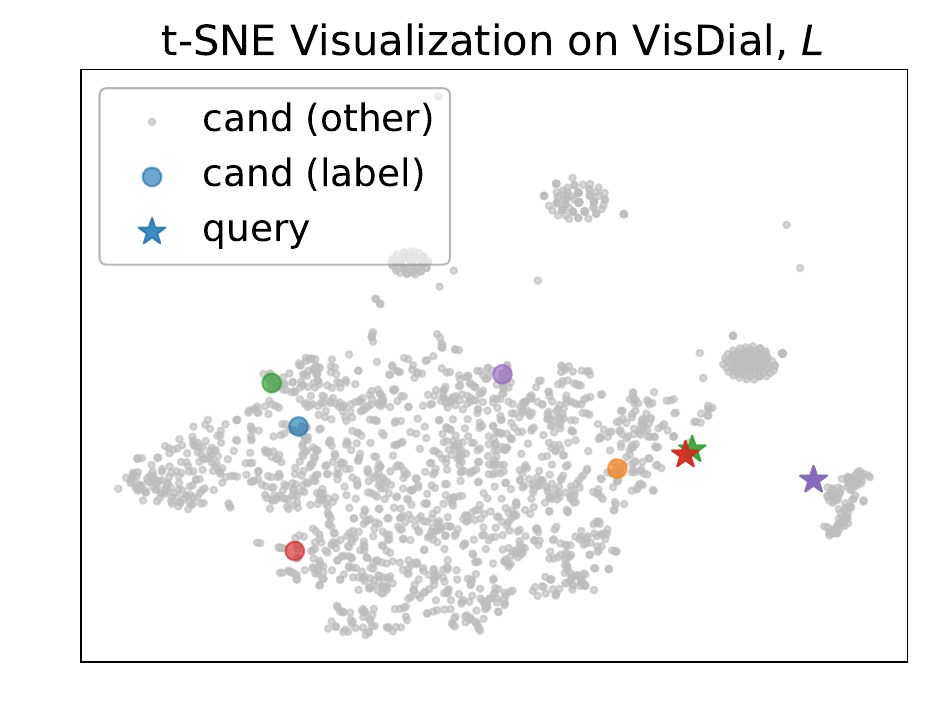}
    \caption{Before training}
  \end{subfigure}

  \vspace{0.5em}

  \begin{subfigure}{\columnwidth}
    \centering
    \includegraphics[width=0.49\columnwidth]{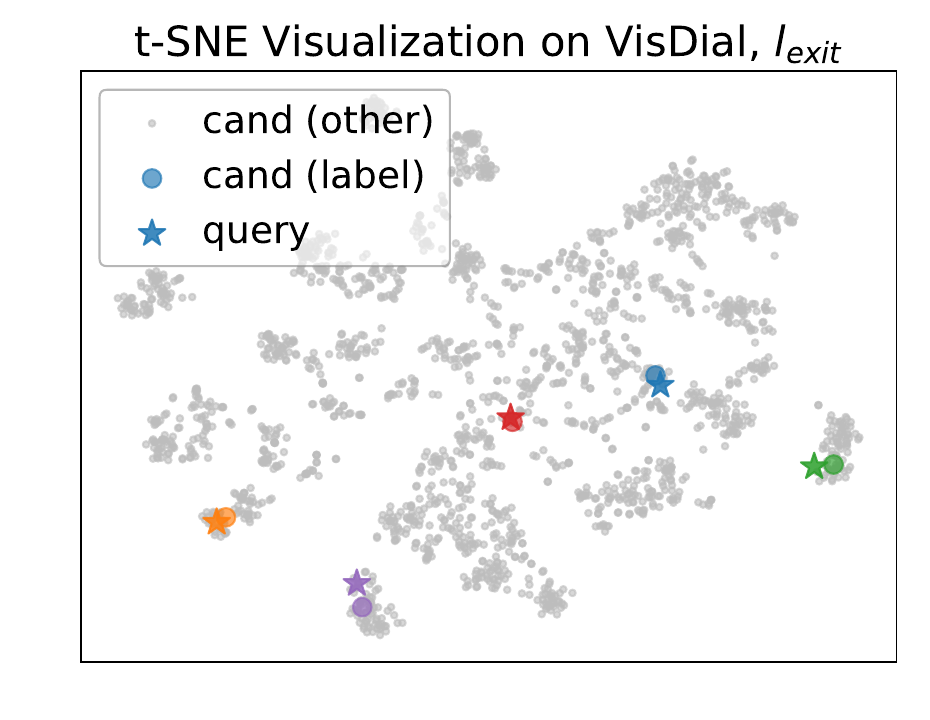}
    \hfill
    \includegraphics[width=0.49\columnwidth]{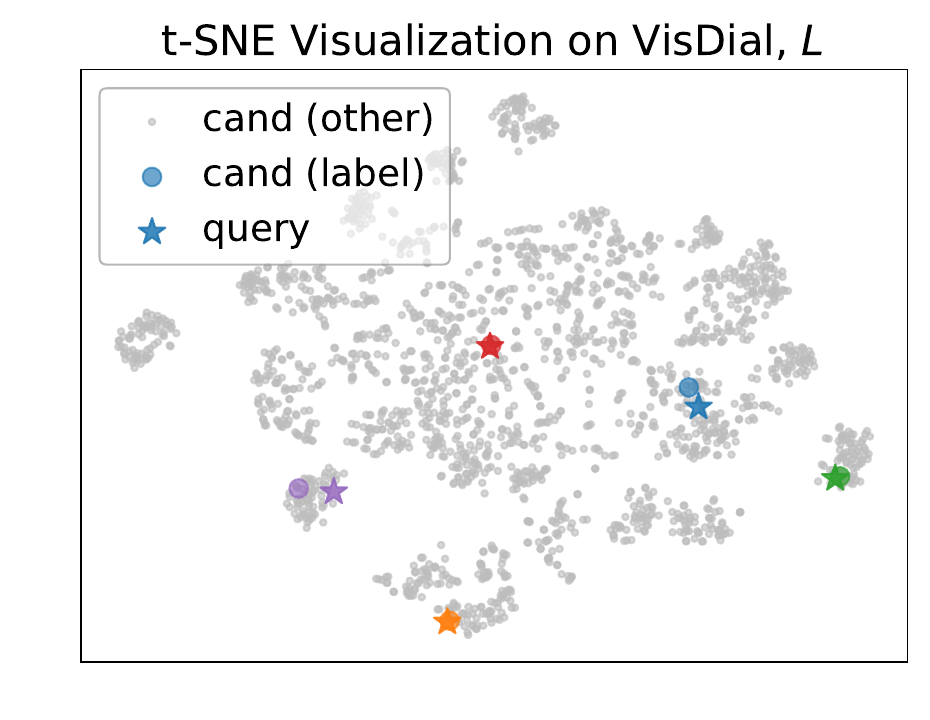}
    \caption{After training}
  \end{subfigure}

  \caption{t-SNE visualizations on VisDial}
\end{figure}

\begin{figure}[htbp]
  \centering

  \begin{subfigure}{\columnwidth}
    \centering
    \includegraphics[width=0.49\columnwidth]{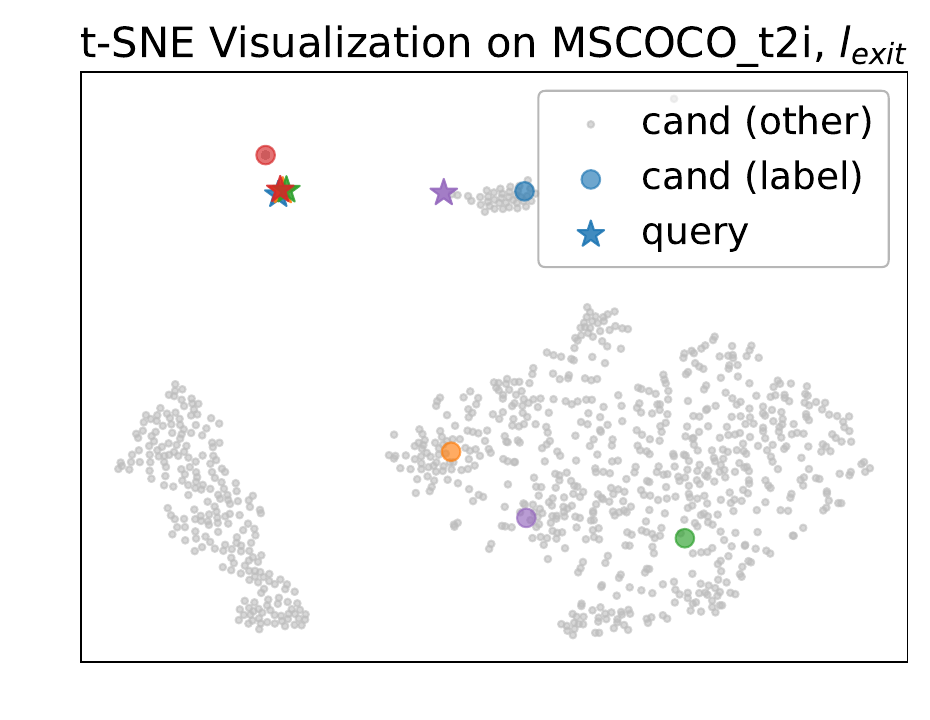}
    \hfill
    \includegraphics[width=0.49\columnwidth]{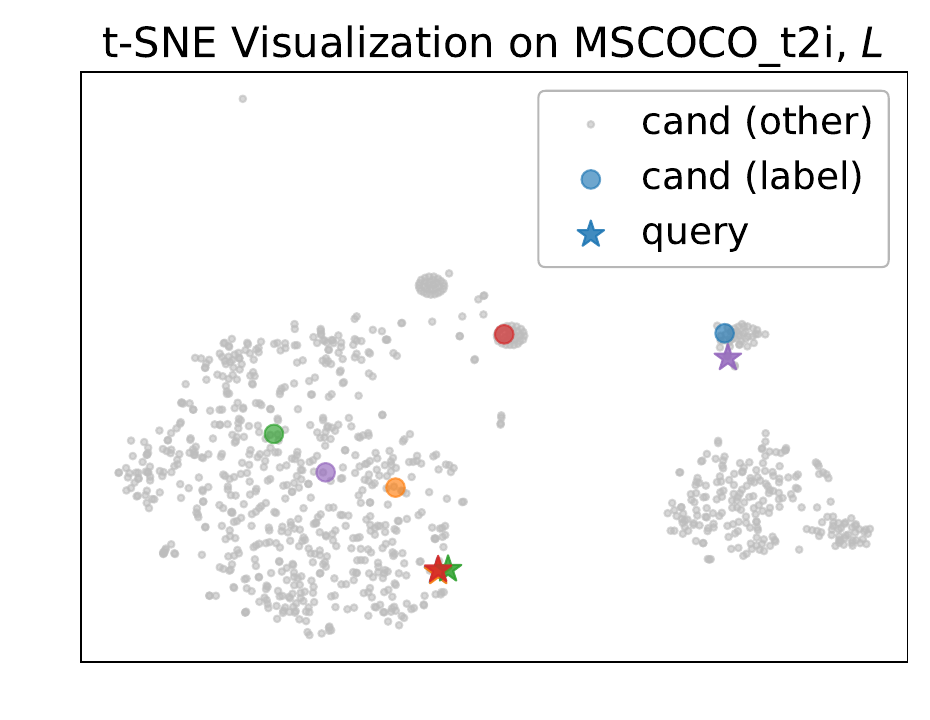}
    \caption{Before training}
  \end{subfigure}

  \vspace{0.5em}

  \begin{subfigure}{\columnwidth}
    \centering
    \includegraphics[width=0.49\columnwidth]{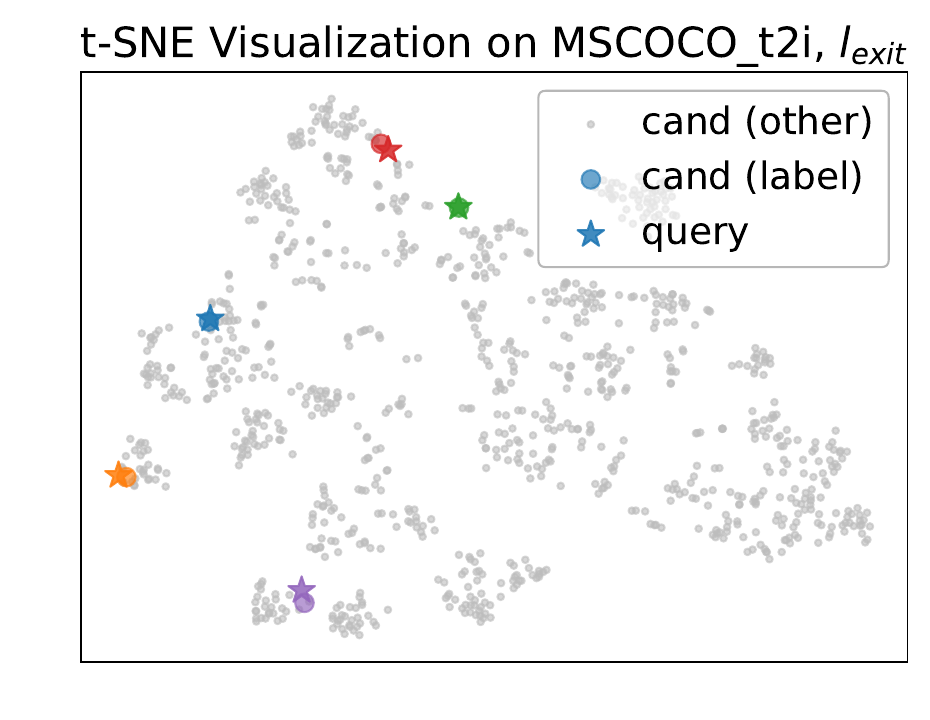}
    \hfill
    \includegraphics[width=0.49\columnwidth]{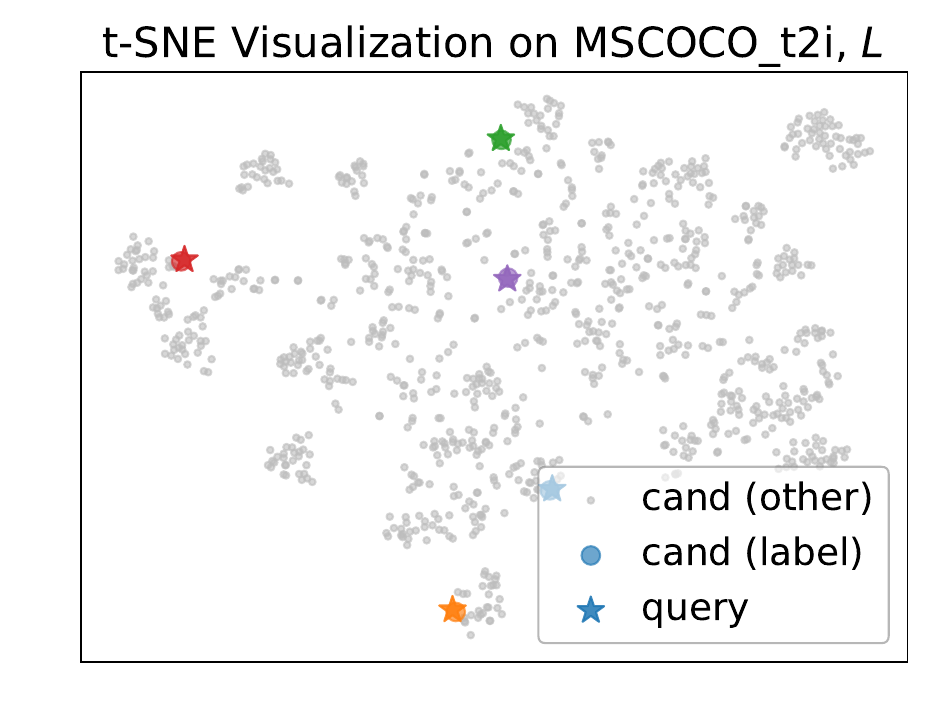}
    \caption{After training}
  \end{subfigure}

  \caption{t-SNE visualizations on MSCOCO\_t2i}
\end{figure}

\begin{figure}[htbp]
  \centering

  \begin{subfigure}{\columnwidth}
    \centering
    \includegraphics[width=0.49\columnwidth]{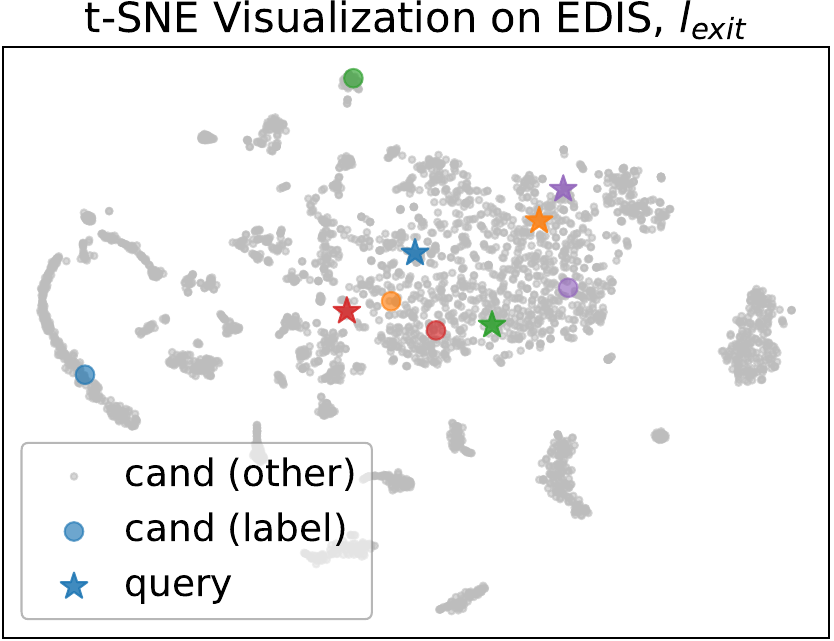}
    \hfill
    \includegraphics[width=0.49\columnwidth]{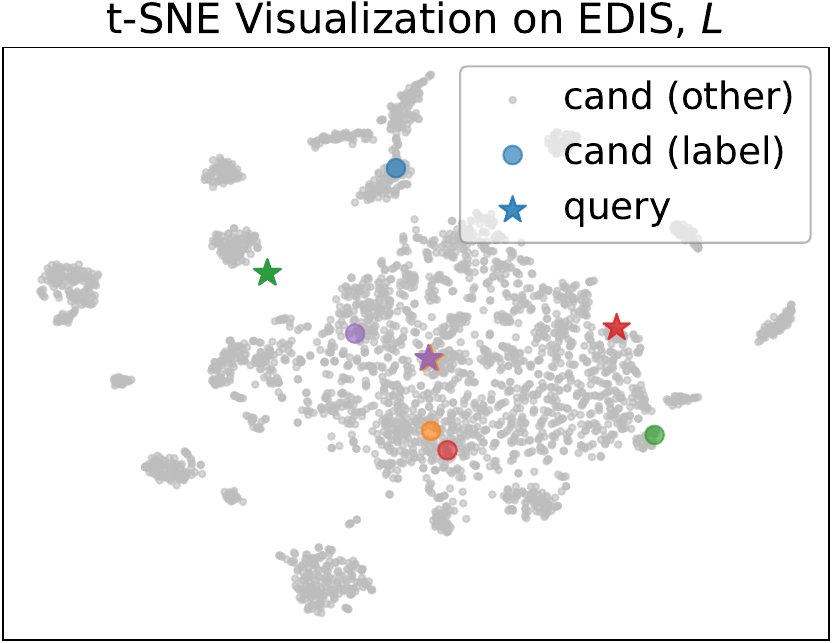}
    \caption{Before training}
  \end{subfigure}

  \vspace{0.5em}

  \begin{subfigure}{\columnwidth}
    \centering
    \includegraphics[width=0.49\columnwidth]{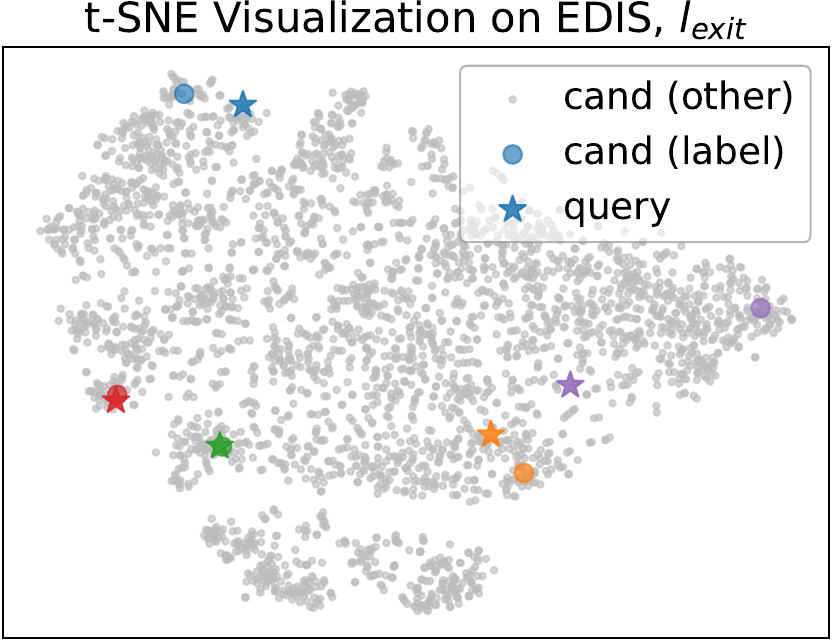}
    \hfill
    \includegraphics[width=0.49\columnwidth]{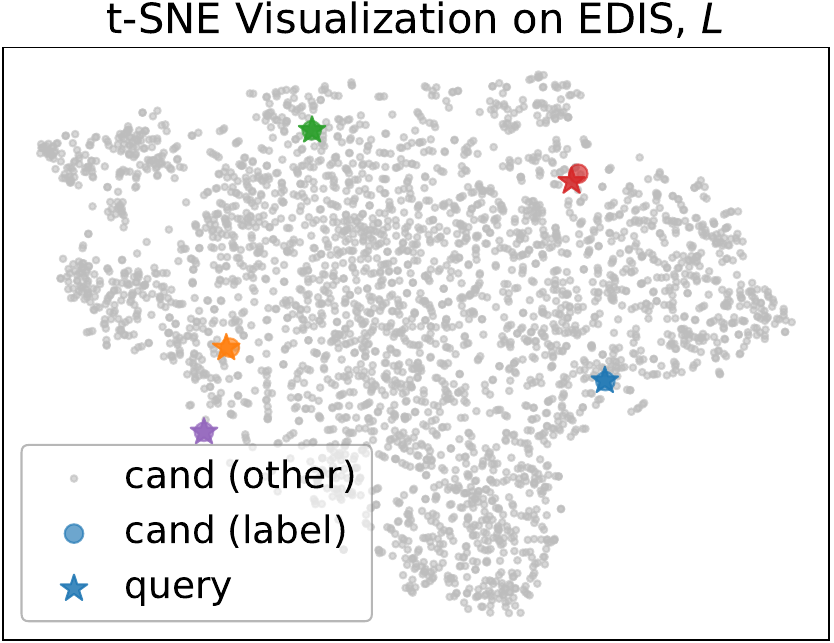}
    \caption{After training}
  \end{subfigure}

  \caption{t-SNE visualizations on Wiki-SS-NQ}
\end{figure}

\begin{figure}[htbp]
  \centering

  \begin{subfigure}{\columnwidth}
    \centering
    \includegraphics[width=0.49\columnwidth]{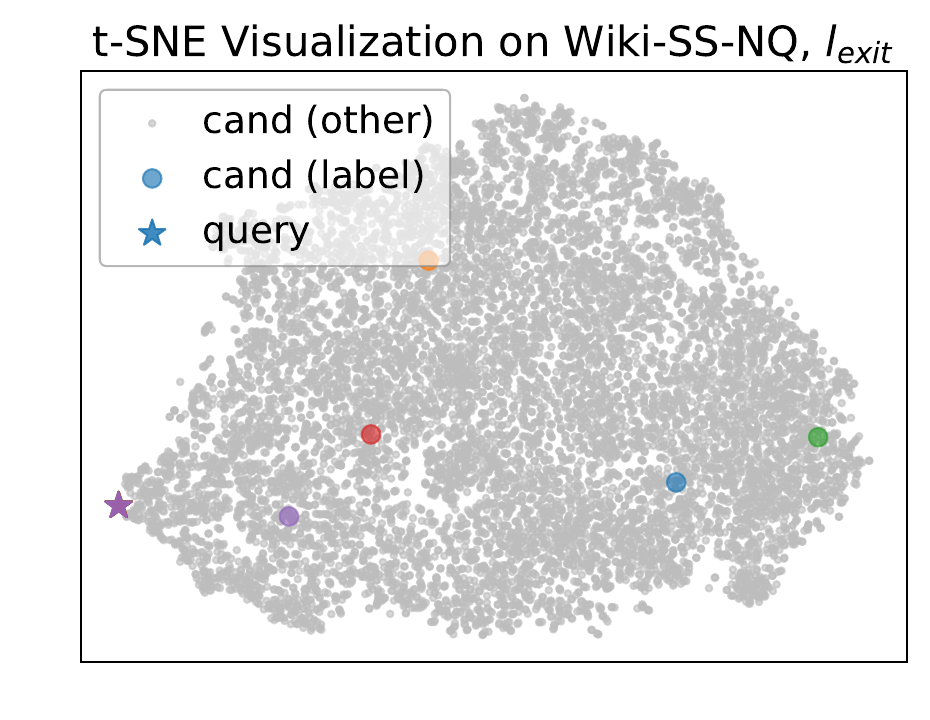}
    \hfill
    \includegraphics[width=0.49\columnwidth]{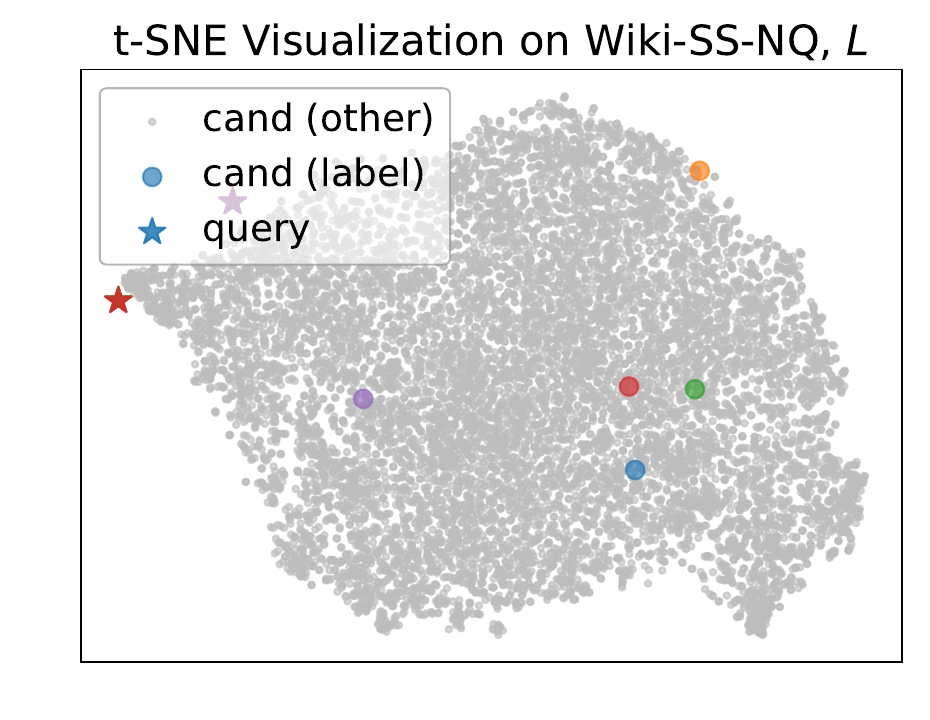}
    \caption{Before training}
  \end{subfigure}

  \vspace{0.5em}

  \begin{subfigure}{\columnwidth}
    \centering
    \includegraphics[width=0.49\columnwidth]{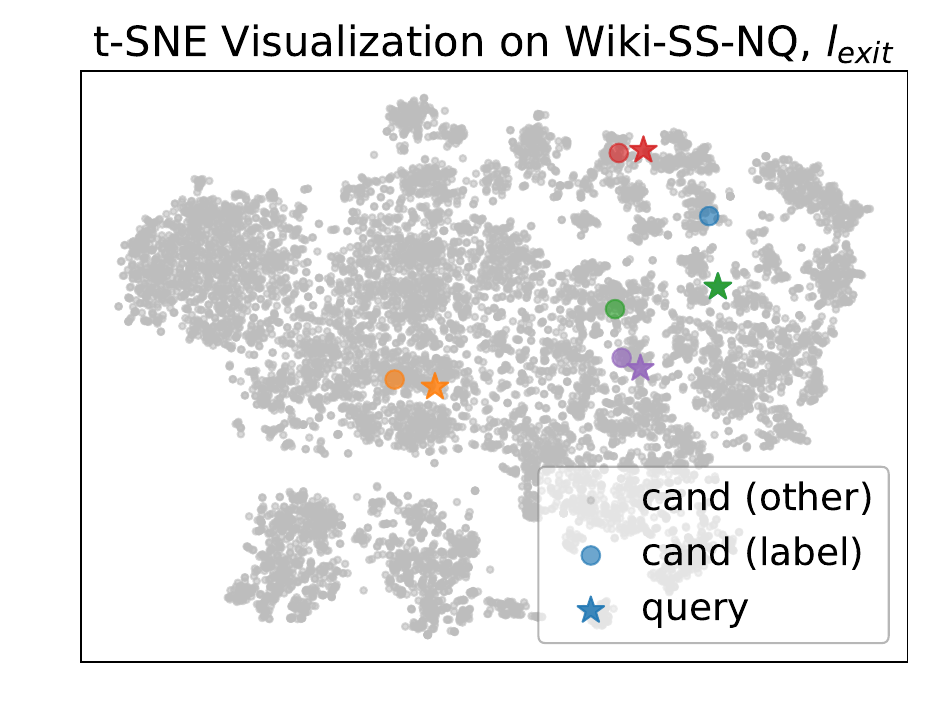}
    \hfill
    \includegraphics[width=0.49\columnwidth]{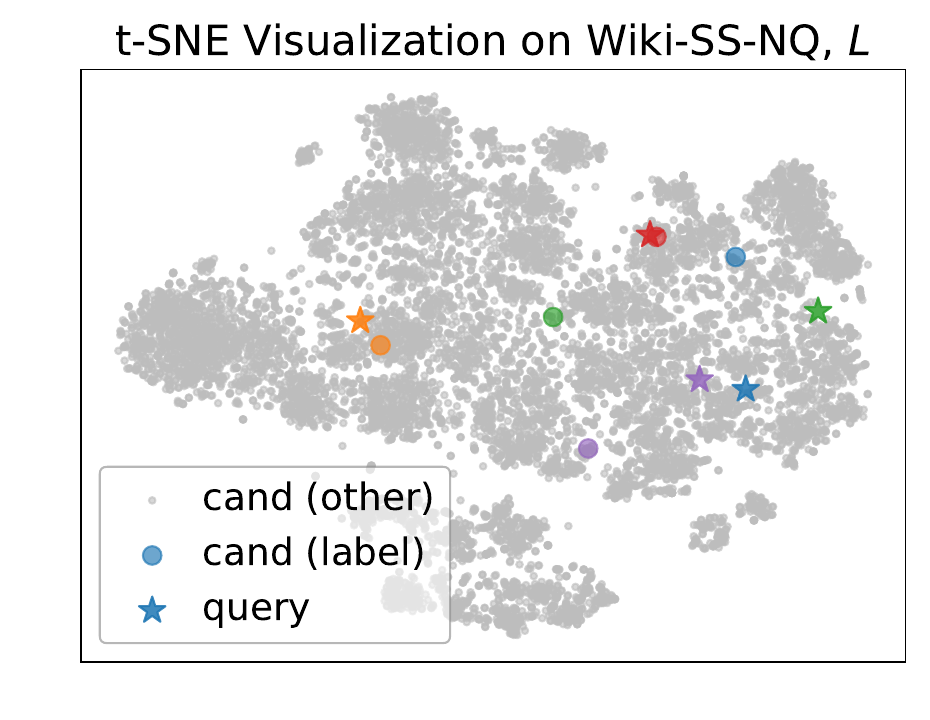}
    \caption{After training}
  \end{subfigure}

  \caption{t-SNE visualizations on Wiki-SS-NQ}
\end{figure}
\label{sec:appendix}
\end{document}